\documentclass[twocolumn]{aastex63}
\usepackage{graphicx}
\usepackage{amssymb}
\usepackage{amsmath}
\usepackage{multirow}
\usepackage{hyperref}
\usepackage{natbib}
\usepackage{comment}
\usepackage{systeme}
\usepackage{footmisc}
\usepackage{booktabs}

\shorttitle{Polarization study of CIZA\,J2242.8+5301}
\shortauthors{G. Di Gennaro et al.}

\begin{document}


\title{Downstream depolarization in the Sausage relic: a 1--4 GHz Very Large Array study}
\author[0000-0002-8648-8507]{G. Di Gennaro}
\affil{Leiden Observatory, Leiden University, PO Box 9513, 2300 RA Leiden, The Netherlands}
\affil{Center for Astrophysics $\mid$ Harvard \& Smithsonian, 60 Garden St., Cambridge, MA 02138, USA}

\author[0000-0002-0587-1660]{R.J. van Weeren}
\affil{Leiden Observatory, Leiden University, PO Box 9513, 2300 RA Leiden, The Netherlands}
\affil{Center for Astrophysics $\mid$ Harvard \& Smithsonian, 60 Garden St., Cambridge, MA 02138, USA}

\author[0000-0001-5636-7213]{L. Rudnick}
\affil{Minnesota Institute for Astrophysics, University of Minnesota, 116 Church St. S.E., Minneapolis, MN 55455, USA}

\author{M. Hoeft}
\affil{Th{\"u}ringer Landessternwarte, Sternwarte 5, 07778 Tautenburg, Germany}

\author[0000-0002-3369-7735]{M. Br{\"u}ggen}
\affil{Hamburger Sternwarte, Universit{\"a}t Hamburg, Gojenbergsweg 112, 21029 Hamburg, Germany}

\author[0000-0002-5455-2957]{D. Ryu}
\affil{Department of Physics, School of Natural Sciences UNIST, Ulsan 44919, Korea}

\author[0000-0001-8887-2257]{H.J.A. R{\"o}ttgering}
\affil{Leiden Observatory, Leiden University, PO Box 9513, 2300 RA Leiden, The Netherlands}

\author[0000-0002-9478-1682]{W. Forman}
\affil{Center for Astrophysics $\mid$ Harvard \& Smithsonian, 60 Garden St., Cambridge, MA 02138, USA}

\author[0000-0001-8322-4162]{A. Stroe}
\altaffiliation{Clay Fellow}
\affil{Center for Astrophysics $\mid$ Harvard \& Smithsonian, 60 Garden St., Cambridge, MA 02138, USA}

\author[0000-0001-5648-9069]{T.W. Shimwell}
\affil{Leiden Observatory, Leiden University, PO Box 9513, 2300 RA Leiden, The Netherlands}
\affil{ASTRON, The Netherlands Institute for Radio Astronomy, Postbus 2, 7990 AA, Dwingeloo, The Netherlands}

\author[0000-0002-0765-0511]{R.P. Kraft}
\affil{Center for Astrophysics $\mid$ Harvard \& Smithsonian, 60 Garden St., Cambridge, MA 02138, USA}

\author[0000-0003-2206-4243]{C. Jones}
\affil{Center for Astrophysics $\mid$ Harvard \& Smithsonian, 60 Garden St., Cambridge, MA 02138, USA}

\author[0000-0002-8286-646X]{D.N. Hoang}
\affil{Hamburger Sternwarte, Universit{\"a}t Hamburg, Gojenbergsweg 112, 21029 Hamburg, Germany}

\correspondingauthor{Gabriella Di Gennaro}
\email{digennaro@strw.leidenuniv.nl}

\received{17 June 2020}
\accepted{11 February 2021}

\begin{abstract}
Radio relics are elongated sources related to shocks driven by galaxy cluster merger events. Although these objects are highly polarized at GHz frequencies ($\gtrsim 20\%$), high-resolution studies of their polarization properties are still lacking. We present the first high-resolution and high-sensitivity polarimetry study of the merging galaxy cluster CIZA\,J2242.8+5301 in the 1--4 GHz frequency band. We use the $QU$-fitting approach to model the Stokes $I$, $Q$ and $U$ emission, obtaining best-fit intrinsic polarization fraction ($p_0$), intrinsic polarization angle ($\chi_0$), Rotation Measure (RM) and wavelength-dependent depolarization ($\sigma_{\rm RM}$) maps of the cluster. Our analysis focuses on the northern relic (RN). For the first time in a radio relic, we observe a decreasing polarization fraction in the downstream region. 
Our findings are possibly explained by geometrical projections and/or by decreasing of the magnetic field anisotropy towards the cluster center. 
From the amount of depolarization of the only detected background radio galaxy, we estimate a turbulent magnetic field strength of $B_{\rm turb}\sim5.6~\mu$Gauss in the relic. Finally, we observe Rotation Measure fluctuations of about 30 rad m$^{-2}$ around at the median value of 140.8 rad m$^{-2}$ at the relic position. 
\end{abstract}


\section{Introduction} \label{sec:intro}
Radio relics are synchrotron sources generally located in the outskirts of merging galaxy clusters. They are elongated, often arc-shaped, and not associated with any optical counterparts.
It is now accepted that these sources trace particles (re)accelerated due to the propagation of shock waves generated by a cluster-cluster merger event \citep[see][for a theoretical and observational review]{brunetti+jones14,vanweeren+19}.
Being synchrotron sources, radio relics are also tracers of the magnetic field in cluster outskirts. Numerical simulations \citep[e.g.][]{dolag+99,bruggen+05,vazza+18}, as well as observations \citep[e.g.][]{govoni+04,bonafede+10a}, show that the magnetic field intensity declines with radius (and hence with particle density) in clusters, with central values of a few $\rm \mu Gauss$ \citep{bonafede+10a}.
On the other hand, it is expected that, during a cluster merger, the un-ordered magnetic fields in the intracluster medium (ICM) are compressed, amplified and aligned with the propagating shock plane, generating strongly linearly polarized emission \citep[$\gtrsim 20\%$, see][]{ensslin+98}. The exact mechanism leading to magnetic field amplification at shocks is not completely understood \citep[see][for a recent review]{donnert+18}. For the typical low Mach numbers of cluster merger shocks ($\mathcal{M}=1-3$), the amplification factor appears to be too small to explain the magnetic field strength measured in relics simply via shock compression, as it is for supernovae remnants \citep[][]{iapichino+bruggen12,donnert+17}. Recently, new high-resolution (i.e., 32~kpc) numerical simulations by \cite{wittor+19} show that the polarized emission from relics should strongly depend on the properties of the upstream magnetic field, with laminar gas flow generating parallel alignment of the electric vectors.
Determining the polarization properties of radio relics thus plays a crucial role in the understanding of these sources, as well as the properties of the ICM.

While studies of magnetic fields of radio galaxies, in the field and in galaxy clusters, have been performed \citep[e.g.][]{bicknell+90,govoni+06,o'sullivan+12,o'sullivan+18,bonafede+10b,frick+11,farnsworth+11,orru+15}, very little information is known on the magnetic field structure in radio relics, with few observational studies performed so far \citep[][]{bonafede+10a,vanweeren+10,vanweeren+12,bonafede+13,ozawa+15,pearce+17,stuardi+19}. 
In this paper, we present a detailed polarization analysis,  performed with the Jansky Very Large Telescope (VLA), of the well-studied merging galaxy cluster CIZA\,J2242.8+5301 (hereafter CIZAJ2242) at $z=0.192$ \citep{kocevski+07}. 

The cluster is the result of the collision of two equal-mass sub-clusters \citep{dawson+15,jee+15}, with a small inclination of the merger axis to the plane of the sky \citep[i.e. $|i|\lesssim 10^\circ$,][]{vanweeren+11}. The cluster hosts two main radio relics, in the north and in the south, several tailed radio galaxies 
and several patches of diffuse emission \citep[see][]{digennaro+18}. High-frequency studies, up to 30~GHz, showed a possible steepening in the integrated radio spectrum\footnote{The radio spectrum is defined as $S_\nu\propto\nu^\alpha$, with $\alpha$ the spectral index.} from $\sim-1.0$ to $\sim-1.6$ at $\nu>2.5$ GHz \citep{stroe+16}, in contrast with the simple picture of a single power-law spectrum predicted from the standard acceleration model \citep[i.e. diffusive shock acceleration, DSA;][]{ensslin+98}. Possible explanations were given by \cite{kang+ryu16}, who suggested a model where a shock passed through a region containing fossil electrons, by \cite{donnert+16}, who suggested the presence of exponential magnetic field amplification in the downstream region (being the shock located at the outermost edge of the relic), 
and by \cite{basu+16}, who proposed a non-negligible contribution from the Sunyaev-Zel'dovich (SZ) effect \citep[also supported by single-dish observations, see][]{loi+17}. 
Single-dish observations revealed that this relic is strongly polarized \citep[up to 60\% at 8.35 GHz,][]{kierdorf+17}, although the poor resolution (i.e. $90''$) strongly limited their analysis. From the relic width (55 kpc) and X-ray downtream velocity (about 1000 km s$^{-1}$), \cite{vanweeren+10} estimated magnetic field strengths of 5 or 1.2 $\mu$Gauss. 

The paper is organized as follows: in Sect. \ref{sec:obs} we describe the data reduction and the imaging procedures; in Sect. \ref{sec:lambda-fit} we present the $QU$ fitting approach; we highlight the effect of the Galactic Rotation Measure in Sect. \ref{sec:galacticRM}; the results and discussion are given in Sect. \ref{sec:res} and \ref{sec:disc}; we end with the conclusion in Sect. \ref{sec:conc}. Throughout the paper, we assume a flat $\Lambda$CDM cosmology with $H_0=70$ km s$^{-1}$ Mpc$^{-1}$, $\Omega_{\rm m}=0.3$ and $\Omega_\Lambda=0.7$, which gives a conversion factor of 3.22 kpc/$^{\prime\prime}$ and a luminosity distance of $\approx944$ Mpc, at the cluster's redshift \citep[$z =0.192$,][]{kocevski+07}.

\section{Observations and data reduction}\label{sec:obs}
We made use of the same 1--4 GHz VLA observations presented in \cite{digennaro+18}, to which we refer for a detailed description of the data reduction. The observations were made with all the four array configurations (namely, A, B, C and D), some of them split into sub-datasets 
\citep[see Table~1 in][]{digennaro+18}. 
Due to the large angular size of the cluster, and the limited field of view (FOV) at 2--4 GHz, we observed three separate pointings in this frequency range. We briefly summarize the data reduction strategy below. 

\begin{table*}
\caption{Datacube information. Columns 1 to 3: Gaussian uv-taper, weighting and robust parameters for the imaging. Column 4: final resolution of the datacubes. Column 5: total number of channels in the 1--2 and 2--4 GHz bands. Column 6: channel width in MHz in the 1--2 and 2--4 GHz bands. Column 7: noise map for the Stokes $I$, $Q$ and $U$ datacubes.}
\vspace{-5mm}
\begin{center}
\begin{tabular}{cccccccccccccc}
\hline
\hline
uv-taper &  weighting	& robust	& resolution & \multicolumn{2}{c}{\#channels} & \multicolumn{2}{c}{$\Delta\nu$} & \multicolumn{3}{c}{$\sigma_{\rm rms[1.26-3.60GHz]}$} \\
$[^{\prime\prime}]$& & & $[^{\prime\prime}\times^{\prime\prime}]$& & & \multicolumn{2}{c}{[MHz]}	& \multicolumn{3}{c}{[$\mu{\rm Jy~beam}^{-1}$]} \\
\hline
 &  &  &  & 1--2 GHz & 2--4 GHz & 1--2 GHz & 2--4 GHz & $I$ & $Q$ & $U$ \\

\cmidrule(lr){5-6}\cmidrule(lr){7-8}\cmidrule(lr){9-11}
2.5 & uniform		& N/A	&  $2.7\times2.7$   & 104 & 75 & 4 & 16 &  12.1 &  11.2 &  11.3 \\
2.5 & Briggs		& 0		&  $4.55\times4.55$ & 104 & 75 & 4 & 16 &  8.9 & 10.1 & 10.0 \\
5   & Briggs		& 0		&  $7\times7$       & 104 & 136 & 4 & 8 & 7.9 & 5.1 & 5.2 \\
10  & Briggs  		& 0		& $13\times13$      & 104 & 136 & 4 & 8 & 18.2 & 5.1 & 5.4 \\
\hline
\end{tabular}
\end{center}
{Note: The noise levels in the last column have been calculated as standard deviation of the datacube, in a central, ``empty'' region of the cluster. For the $2.5^{\prime\prime}$-tapered images, we only produced stamps of the single sources, hence we report the map noise locally to RN.}		
\label{tab:wsclean}
\end{table*}

First, we Hanning smoothed the data, and removed radio frequency interference (RFI) with the {\it tfcrop} mode from the \texttt{flagdata} task in \texttt{CASA}. Then, we calibrated the antenna delays, bandpass, cross-hand delays, and polarization leakage and angles using the primary calibrators 3C138, 3C147, and/or 3C48. For the polarization leakage calibration, we can only make use of an unpolarized source\footnote{In principle, a calibrator with enough parallactic angle coverage can also be used for the leakage calibration. This kind of calibrator was not available in our observations.}, hence we discarded all the sub-datasets where 3C48 was the only calibrator \citep[for further details, see][]{digennaro+18}. We determined the global cross-hand delay solutions (\texttt{gaintype=`KCROSS'}) from the polarized calibrator 3C138, taking a RL-phase difference of  $-10^\circ$ (both L- and S-band) and polarization fractions of 7.5\% and 10.7\% (L- and S-band respectively). We used 3C147 to calibrate the polarization leakage terms (\texttt{poltype=`Df'}), and 3C138 to calibrate the polarization angle (\texttt{poltype=`Xf'}).
The solution tables were applied on the fly to determine the complex gain solution for the secondary calibrator J2202+4216. Additional RFI removal was performed, using the {\it tfcrop} and {\it rflag} modes (in \texttt{CASA}) and \texttt{AOFlagger} \citep{offringa+10}, before and after applying the calibration tables to the target field, respectively. The data were averaged by a factor of two in time and a factor of four in frequency.
This reflects a frequency resolution (i.e. channel width) of $\Delta\nu=4$ and $\Delta\nu=8$ MHz, at 1--2 and 2--4 GHz respectively. The only exception is the $2.5^{\prime\prime}$-tapered dataset at 2--4 GHz, for which we average by a factor of eight, i.e. $\Delta\nu=16$ MHz.
Finally, self-calibration was performed to refine the amplitude and phase calibration on the target. 

To retrieve the images for all the Stokes parameters (i.e., $I$, $Q$ and $U$) at each channel $\Delta\nu$, as required for a detailed polarization analysis, we employed the \texttt{WSClean} \citep{offringa+14}. 
Images were produced with different weightings (i.e. \texttt{Briggs} and \texttt{uniform}), and uv-tapers (i.e., $2.5^{\prime\prime}$, $5^{\prime\prime}$ and $10^{\prime\prime}$). Bad spectral windows and channels were discarded from the final analysis. 
For the Stokes-$Q$ and -$U$ images, we also used the options \texttt{-join-channels}, \texttt{-join-polarizations} and \texttt{-squared-channel-joining}, which prevent the $Q$-, $U$-flux to be averaged out to zero\footnote{\url{https://sourceforge.net/p/wsclean/wiki/RMSynthesis/}}. After imaging, channel images that where too noisy or low-quality were removed. In the end, a total of 240 channels, for the $5''$- and $10''$-tapered images, and 179 channels, for the $2.5''$-tapered images, were used. This results in a final frequency coverage of 1.26--3.60 GHz.
The single-channel images were re-gridded to the same pixel grid and convolved to the same resolution (see Tab \ref{tab:wsclean}). Finally, all the single images were primary-beam corrected, by taking the beam variation with the frequency taken into account\footnote{The beam shapes have been obtained with \texttt{CASA} v. 5.3.}, and merged into a single datacube for each Stokes parameter. Errors in the single channel images were estimated using the rms noise level from a central, empty, region of the cluster (at $7''$ and $13''$ resolution) or locally for the sources of interest (at $4.5''$ and $2.7''$ resolution).

\section{Polarization theory and modelling approach}\label{sec:lambda-fit}
The linear polarization emission can be described in terms of Stokes parameters for the total intensity, $I$, and the orthogonal components, $Q$ and $U$:
\begin{equation}\label{eq:pol_stokes}
P(\lambda^2) = p(\lambda^2)I(\lambda^2)\exp[2i\chi(\lambda^2)] = Q(\lambda^2) + iU(\lambda^2)
\, ,
\end{equation}
and $\lambda$ is the observing wavelength. Here, $p(\lambda^2)$ is the fractional (or degree of) polarization and $\chi(\lambda^2)$ is the polarization angle, which are wavelength-dependent quantities that can be written as:

\begin{equation}\label{eq:frac_pol}
p(\lambda^2) = \frac{P(\lambda^2)}{I(\lambda^2)} = \frac{\sqrt{Q^2(\lambda^2) + U^2(\lambda^2)}}{I(\lambda^2)}
\end{equation}
and
\begin{equation}\label{eq:ang_pol}
\chi(\lambda^2) = \frac{1}{2}\arctan \left (\frac{U(\lambda^2)}{Q(\lambda^2)} \right ) \, .
\end{equation}

The passage of the polarized radiation through a foreground magneto-ionic medium, such as the ICM, results in a rotation of polarization plane via the Faraday effect according to
\begin{equation}\label{eq:chi_lambda2}
\chi(\lambda^2) = \chi_0 + {\rm RM}\lambda^2 \, ,
\end{equation}
where $\chi_0$ is the intrinsic polarization angle and RM is the Faraday rotation measure. This is defined as:

\begin{figure*}
\centering
{\includegraphics[width=\textwidth]{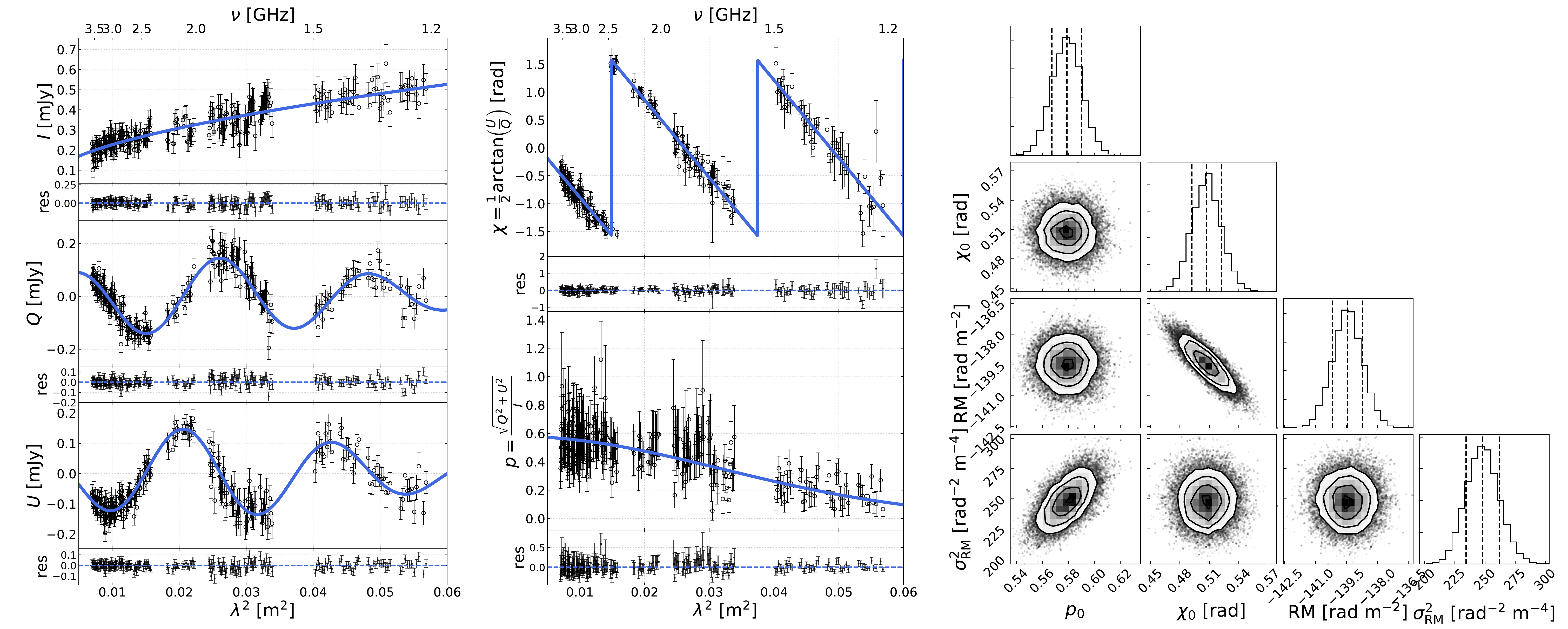}}
\caption{Result of the $QU$ fit assuming the external depolarization model (EDF, Eq. \ref{eq:eq:ext_far}) on a single pixel of the northern relic. Left panel: Fits on Stokes $I$, $Q$ and $U$ fluxes. 
Central panel: Resulting fractional polarization, $p(\lambda^2)$, and polarization angle, $\chi(\lambda^2)$, estimated from Eqs. \ref{eq:frac_pol} and \ref{eq:ang_pol}, respectively. Right panel: Corner plot for the distribution of the uncertainties in the fitted polarization parameters (i.e. $p_0$, $\chi_0$, RM and $\sigma_{\rm RM}^2$); contour levels are drawn at $[0.5,1.0,1.5,2.0]\sigma$, with $\sigma$ the 68\% statistical uncertainty (see dashed lines in the 1D histogram). 
}
\label{fig:fitpixelexample}
\end{figure*}

\begin{equation}\label{eq:rm_fit}
{\rm RM} = 
0.81 \int_{\rm source}^{\rm observer} n_e B_\parallel dl \quad {\rm [rad~ m^{-2}]} \, ,
\end{equation}
where $n_e$ is the electron density (in cm$^{-3}$), $B_\parallel$ the magnetic field (in $\mu$Gauss) along the line of sight, $l$ the path length through the magneto-ionic medium (in pc), and with the sign of the equation defined positive for a magnetic field pointing towards the observer.

The traditional way to retrieve the intrinsic polarization angle $\chi_0$ is to observe $\chi$ at several wavelengths, and linearly fit Eq. \ref{eq:chi_lambda2}. 
The long-standing problem of this approach is the lack of a sufficient number of $\chi(\lambda^2)$ measurements. In this work, this issue is overcome by the large number of channel images with high signal-to-noise (S/N) of our wide-band observations (see Sect. \ref{sec:QUmod}).

Several models of the polarized signal, in the presence of Faraday rotation, are known. In the simplest scenario, Eq. \ref{eq:pol_stokes} can be written as:
\begin{equation}\label{eq:far_rot}
P(\lambda^2) = p_0 I \exp[2i(\chi_0 + {\rm RM}\lambda^2)] \, ,
\end{equation}
with $p_0$ the intrinsic polarization fraction. This corresponds to the physical situation of a single Faraday screen in the foreground.  
In this case, $d\chi/d\lambda^2$ and $p(\lambda)$ are constant.

Observations have shown that radio relics depolarize at frequencies $\lesssim 1$ GHz \citep{brentjens11,pizzo+11,ozawa+15}. Common depolarization mechanisms are external and internal Faraday rotation dispersion \citep[EFD and IFD, respectively; see][for the detailed parametrization of those mechanisms]{sokoloff+98}. EFD occurs when variations in the magnetic field direction are not resolved in the single beam \citep{burn66,tribble91}. For a Gaussian distribution of RM, the observed polarization is parameterized as:
\begin{equation}\label{eq:eq:ext_far}
P(\lambda^2) = p_0 I \exp(-2\sigma^2_{\rm RM}\lambda^4) \exp[2i(\chi_0 + {\rm RM}\lambda^2)] \, ,
\end{equation}
where $\sigma_{\rm RM}$ is the dispersion about the mean RM across the beam on the sky.

On the other hand, IFD occurs when the emitting source and the Faraday screen (i.e. the rotating layer) are mixed. In this case, depolarization is due to the random direction of the plane of polarization through the emitting region, and it can be parametrized as:
\begin{equation}\label{eq:eq:int_far}
P(\lambda^2) = p_0 I \left [ \frac{1-\exp(-2\varsigma^2_{\rm RM}\lambda^4)}{2\varsigma^2_{\rm RM}\lambda^4} \right ] \exp[2i(\chi_0 + {\rm RM}\lambda^2)] \, ,
\end{equation}
where $\varsigma_{\rm RM}$ is the internal dispersion of the random field. 

\subsection{{\it QU}-modelling approach}\label{sec:QUmod}
Stokes $Q(\lambda^2)$ and $U(\lambda^2)$ fitting has been used in literature to determine the polarization properties of a magneto-ionic layer \citep[e.g.][]{o'sullivan+12,ozawa+15,anderson+16}. 
In this approach, $Q(\lambda^2)$ and $U(\lambda^2)$ were fitted simultaneously
with cosine and sine models, while $I(\lambda^2)$ was fitted with a log-parabolic model \citep[see also][]{massaro+04}, which represents a curved spectrum, as suggested by \cite{stroe+16} and given the large bandwidth used:

\begin{equation}\label{eq:Ifit}
I_\nu = I_0\nu^{a+b\log(\nu/\nu_{\rm ref})} \, ,
\end{equation}
where we fixed the reference frequency $\nu_{\rm ref}$ to 1 GHz.

In this model, $b$ is the curvature parameter and the spectral index is calculated as the log-derivative, i.e. $\alpha= a+2b\log(\nu/\nu_{\rm ref})$. 
For each channel image in the $I(\lambda^2)$, $Q(\lambda^2)$ and $U(\lambda^2)$ datacubes, the uncertainties were computed by adding in quadrature the relative (spatial) map noise and 5\% of the Stokes $I$, $Q$ and $U$ flux in each channel. Here, the 5\% represents a spatially-independent intrinsic scatter which takes into account the flux variations between the single-frequency channel maps. 
The origin of this scatter is not fully clear, but it is probably related to bandpass calibration and/or deconvolution uncertainties.

We fitted our data with the Markov Chain Monte Carlo (MCMC) method\footnote{The initial guesses for the parameters were obtained with the least square method (\emph{scipy.optimize.leastsq} in Python).} \citep{foreman-mackey+13} to explore the best-set of model parameters \citep{ozawa+15}. During the fitting procedure, all the parameters (i.e. $I_0$, $a$ and $b$ for Stokes $I$, and $p_0$, $\chi_0$, RM and $\sigma^2_{\rm RM}$ for the combined Stokes $Q$ and $U$) were left free to vary through the full parameter space. In the fitting, we constrained $p_0$, $\chi_0$ and $\sigma^2_{\rm RM}$ (or $\varsigma^2_{\rm RM}$) to the following physical conditions:

\begin{equation}
\begin{cases}
0 \leq p_0 \leq 1 \\
0 \leq \chi_0 < \pi
\\
\sigma_{\rm RM}^2 \geq 0 ~{\rm or}~ \varsigma_{\rm RM}^2 \geq 0\, ,
\end{cases}
\end{equation}
and we assumed a single-RM component model (see also Appendix \ref{apx:dwnpix})). The upper limit for the polarization angle is set to $\pi$ because the polarization vectors have no preferred direction. In this convention, $\chi_0=0$ and $\chi_0=\pi/2$ give the north/south and east/west directions, respectively. We chose to include depolarization in our fit as our observations showed a decrease in polarization fraction towards longer $\lambda^2$. It is worth noting that the $p_0$ value obtained from the MCMC fit could be an underestimation of the intrinsic polarization fraction, because of the limited $\lambda^2$ coverage, and
possible misalignment of the intrinsic polarization angle $\chi_0$ from different emitting sites along the line of sight.  
Hereafter, we refer to $p_0$ as the best-fit intrinsic polarization fraction.  
The uncertainties on the best-fitting parameters were determined with the MCMC analysis. The results of the fitting procedure using the EFD model on a representative single pixel in the cluster northern relic are displayed in Fig. \ref{fig:fitpixelexample}. Similar result were found using the IDF model (Eq. \ref{eq:eq:int_far}), except for $\varsigma_{\rm RM}$ which is higher due to the different functional way it describes the depolarization.

\section{Rotation Measure from our Galaxy}\label{sec:galacticRM} 
The best-fit Rotation Measure value obtained could, in principle, give information on the magnetic field structure of the diffuse radio emission in the cluster (Eq. \ref{eq:rm_fit}). However, in order to have a reliable estimation of the RM associated with the ICM, the contribution of the foreground Galactic RM needs to be estimated and removed from the calculations.

\begin{table}
\caption{Averaged RM values of the sources labelled in Fig. \ref{fig:RMcompactsources} observed in the 1--2 GHz frequency range. The ``uncertainty'' on RM is represented by the standard deviation of the RM pixel distribution within the source.}
\vspace{-5mm}
\begin{center}
\begin{tabular}{cccc}
\hline
\hline
Source &  $\rm RA_{J2000}$ & $\rm DEC_{J2000}$ & $\langle {\rm RM} \rangle \pm std({\rm RM})$ \\
& [$^{\rm h~m~s}$] & [$\rm ^\circ~^\prime~^{\prime\prime}$] & [$\rm rad~m^{-2}$]  \\
\hline	
1 & 22 44 31.5 & +53 00 39.0 & $-113.0\pm5.4$  \\
2 & 22 42 12.4 & +52 47 56.5 & $-43.9\pm3.6$  \\
3 & 22 42 05.2 & +52 59 32.0 & $+1.2\pm8.1$  \\
4 & 22 41 22.1 & +53 02 15.5 & $-71.7\pm7.2$  \\ 
5 & 22 41 00.1 & +53 04 15.7 & $-77.4\pm5.1$  \\
6 & 22 41 33.1 & +53 11 07.7 & $-155.9\pm1.4$  \\
7 & 22 43 02.2 & +53 19 42.2 & $-76.0\pm8.5$  \\
8 & 22 43 37.5 & +53 09 15.5 & $-137.2\pm5.0$  \\ 
9 & 22 41 22.9 & +52 52 54.3 & $-81.1\pm6.7$  \\ 
10& 22 43 05.2 & +53 17 33.8 & $-92.0\pm6.4$  \\ 
\hline
\end{tabular}
\end{center}
{Note: 
Source 3 and source 8 are labelled as source A and N in Fig. \ref{fig:pol_emiss}, respectively}
\label{tab:RMcompactsources}
\end{table}

The Galactic coordinates of CIZAJ2242 are $l=104^\circ$ and $b=-5^\circ$, meaning that the cluster lies on close to the Galactic plane. 
Hence, the RMs of the cluster sources are strongly affected by the Faraday rotation from our Galaxy.
Using the map of the Galactic contribution to Faraday rotation provided by \cite{oppermann+15}\footnote{\url{https://wwwmpa.mpa-garching.mpg.de/ift/faraday/2014/index.html}}, we found an average contribution of about $\rm -65\pm57~rad~m^{-2}$ in a region of $20^\prime$ around the cluster center coordinates. 
However, the current available Galactic RM map is affected by very poor angular resolution (i.e. $\rm \sim10^\prime/pixel$), which is comparable with the cluster size ($\sim 15^\prime$). For this reason, we lack detailed information on the RM variations on the cluster/sub-cluster scale. 

\begin{figure*}
\centering
\includegraphics[width=\textwidth]{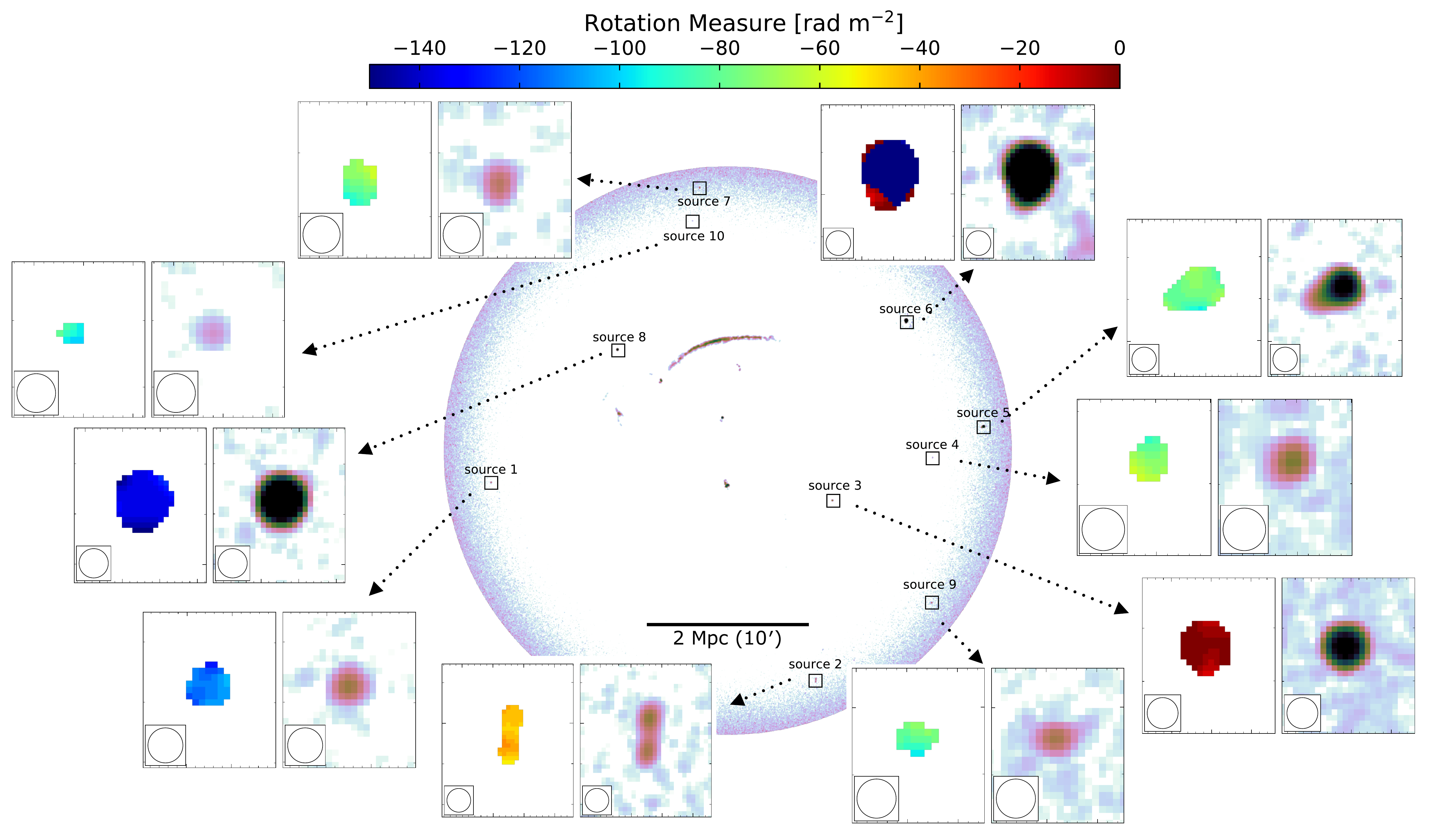}
\caption{Total polarized emission of the 1--2 GHz field of view (${\rm FOV}\sim18'$ in radius) to search for polarized radio galaxies outside CIZAJ2242. A zoom on those sources is shown in the insets, where the Rotation Measure and the total intensity are displayed in the left and right panel, respectively. The RM colorscale is fixed for all the sources. The averaged RM values of those sources are listed in Table~\ref{tab:RMcompactsources}.}
\label{fig:RMcompactsources}
\end{figure*}

We investigated the RM values of compact sources within the field of view of our observations, but outside the cluster region. In this way, we exclude the contribution of the ICM on the RM estimation. Since the size of the primary beam depends on the frequency as $\rm FOV \propto \nu^{-1}$, and we want to maximize the area where we search for polarized sources, we only used the 1--2 GHz observations.  
We found a total of 10 sources in the 1--2 GHz FOV ($\sim18^\prime$, see Fig. \ref{fig:RMcompactsources}). Their Rotation Measure values, listed in Table~\ref{tab:RMcompactsources}, are consistent with the average Galactic RM value found by \cite{oppermann+15}, with a median value of about $-80$ rad m$^{-2}$ and standard deviation of about 42 rad m$^{-2}$. Moreover, we found that sources close to each other (i.e., sources~4 and~5, and sources~7 and~10) have similar RM, suggesting that the Galactic foreground might remain approximately constant in that region, on those spatial scales ($3'-5'$, i.e. few hundreds of kpc, at the cluster distance).
However, we find a strong variation from in RM north to south and east to west, although without a clear trend. It remains therefore difficult to quantify a unique Rotation Measure value from the Galactic foreground, and to subtract it from our measured RM values for the cluster sources. For this reason, in the following maps and plots we report the best-fit RM value, including the Galactic contribution.

\begin{figure*}
\centering
\includegraphics[width=0.9\textwidth]{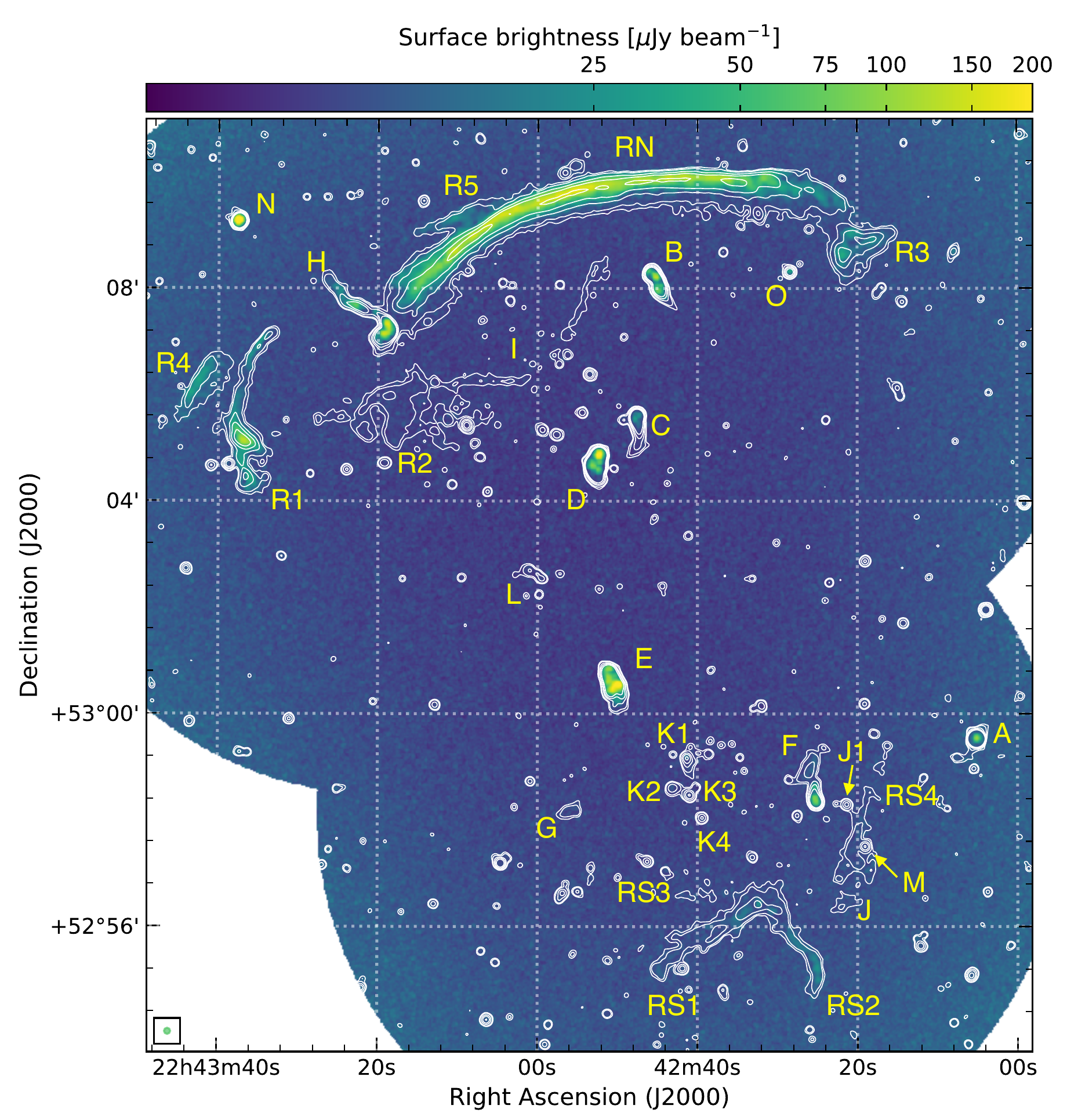}
\caption{Total averaged polarized emission for CIZAJ2242 in the 1.26--3.60 GHz band (effective frequency of 2.3 GHz) at $7^{\prime\prime}$ resolution. This image is not corrected for the Ricean bias. The radio contours are from the averaged total intensity image, in the same frequency band and at the same resolution, with contours drawn at levels of $3\sigma_{\rm rms}\times\sqrt{[1, 4, 16, 64, 256,\ldots]}$, with $\sigma_{\rm rms}=4.2~\mu$Jy~beam$^{-1}$. Sources are labelled following Fig. 2 in \cite{digennaro+18}.}\label{fig:pol_emiss}
\end{figure*}

\begin{figure*}
\centering
{\includegraphics[width=\textwidth]{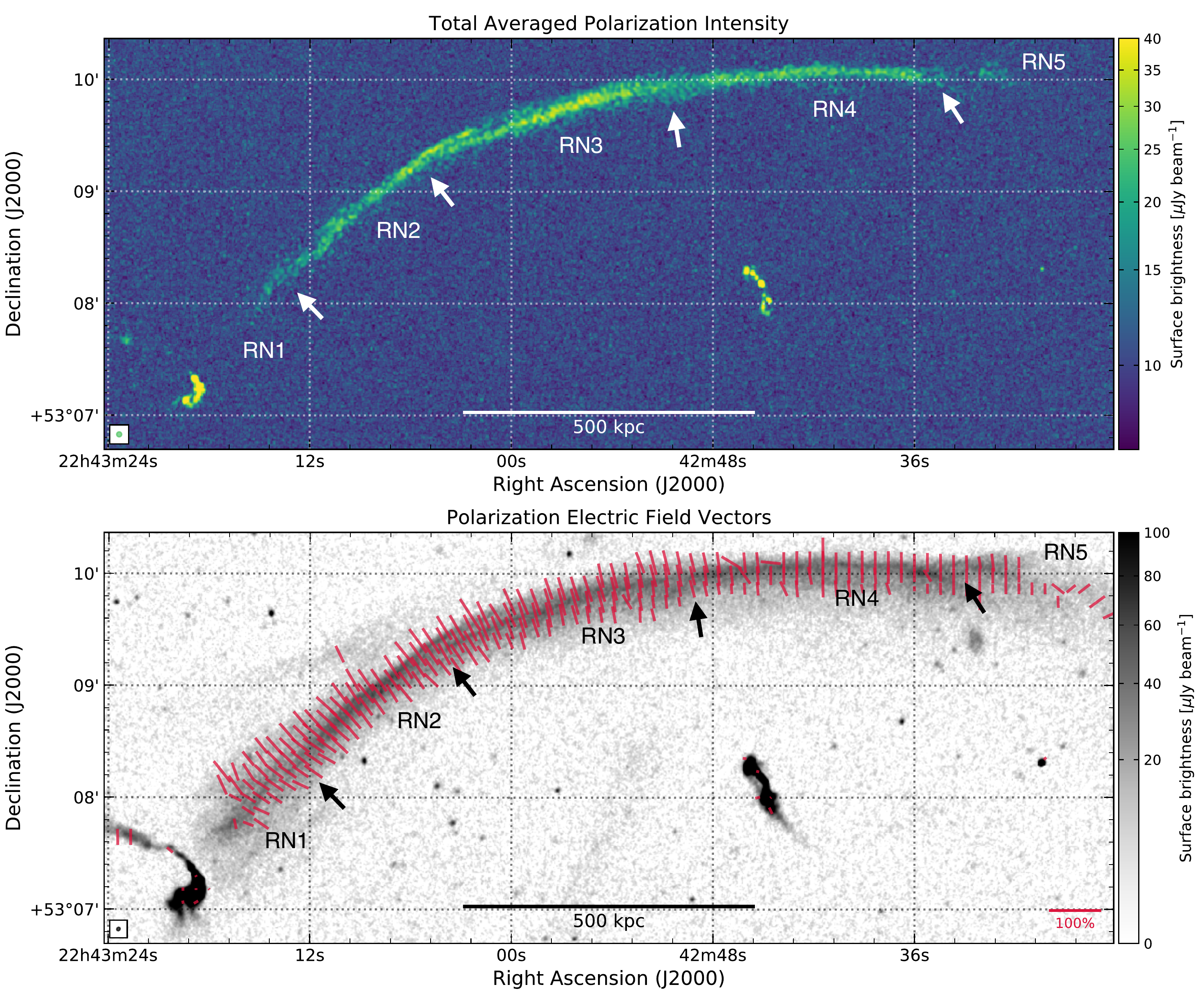}}
\caption{Top panel: High-resolution ($2.7''\times2.7''$) total averaged polarized image in the 1.26--3.60 GHz band (effective frequency of 2.0 GHz) zoomed on the northern relic ($\sigma_{\rm \emph{Q}, rms[1.26-3.60GHz]}=11.2$ and $\sigma_{\rm \emph{U}, rms[1.26-3.60GHz]}=11.3$ $\mu$Jy beam$^{-1}$). As for Fig. \ref{fig:pol_emiss}, this image is not corrected for the Ricean bias. Bottom panel: High-resolution ($2.1''\times1.8''$) Stokes $I$ observation in the 1--2 GHz band \citep{digennaro+18} with the polarization electric field vectors at $2.7''$ resolution, corrected for Faraday Rotation, displayed in red; the length of the vectors is proportional to the intrinsic polarization fraction (scale in the bottom right corner). White and black arrows in the two panels indicate the points where the relic breaks into separate filaments, following Fig. 7 in \cite{digennaro+18}.}
\label{fig:RNzoom}
\end{figure*}

\section{Results}\label{sec:res}
\subsection{Polarized flux densities and fractions}
We obtained the total averaged polarization images in the 1.26--3.60 GHz band by means of the RM-Synthesis technique \citep[][]{brentjens+debruyn05}, using the \texttt{pyrmsynth} tool\footnote{\url{https://github.com/mrbell/pyrmsynth}}. In Fig. \ref{fig:pol_emiss} and in the top panel of Fig. \ref{fig:RNzoom}, we show the total averaged polarization images of the entire cluster at $7''$ resolution and of the northern relic at $2.7''$ resolution, at the effective frequencies of 2.3 and 2.0 GHz, respectively. 
We retrieve the polarized intensity at the canonical frequencies, i.e. 1.5 and 3.0 GHz (i.e. at wavelength of 0.2 and 0.1 m, respectively), using the fit results of Eq. \ref{eq:eq:ext_far} as described in Section \ref{sec:QUmod}. In Table~\ref{tab:fluxes} we report the polarized and total flux densities, the correspondent factional polarization (Eq. \ref{eq:ang_pol}), and the amount of depolarization ${\rm DP^{3.0GHz}_{1.5GHz}}=1-(p_{\rm 1.5GHz}/p_{\rm 3.0GHz})$\footnote{In this convention, ${\rm DP^{3.0GHz}_{1.5GHz}}=0$, i.e. $p_{\rm 1.5GHz}=p_{\rm 3.0GHz}$, means no depolarization, while ${\rm DP^{3.0GHz}_{1.5GHz}}=1$, i.e. $p_{\rm 1.5GHz} \sim 0$, means full depolarization.}, for the diffuse radio sources in the cluster.

\begin{table*}
\caption{Polarized ($P_{\nu}$) and total intensity ($I_{\nu}$) flux densities, and integrated polarization fraction ($p_{\nu}$) for the diffuse radio sources labelled in Fig. \ref{fig:pol_emiss} at $\rm \nu=1.5~and~3.0~GHz$. The depolarization fraction between the two frequencies is shown in the last column.}
\vspace{-5mm}
\begin{center}
\begin{tabular}{cccccccccccccc}
\hline
\hline
Source  &  resolution & $P_{\rm 3.0GHz}^{\rm (a)}$	& $I_{\rm 3.0GHz}$ & $p_{\rm 3.0GHz}^{\rm (b)}$ & $P_{\rm 1.5GHz}^{\rm a}$	& $I_{\rm 1.5GHz}$ & $p_{\rm 1.5GHz}^{\rm (b)}$ & $\rm DP^{3.0GHz}_{1.5GHz}$ \\
        & [$^{\prime\prime}\times^{\prime\prime}$] & [mJy] & [mJy]  &  & [mJy] & [mJy]  \\
\hline	
RN & $7\times7$ & $17.0\pm0.9$ & $45.5\pm2.3$ & $0.37$ & $19.4\pm1.0$ & $105.2\pm5.3$ & $0.18$ & $0.51$\\
RS1+RS2 & $13\times13$ & $1.2\pm0.1$ & $5.7\pm0.3$  & $0.22$ & $2.3\pm0.1$ & $11.2\pm0.6$ & $0.20$ & $0.06$ \\
R1 & $7\times7$ & $1.5\pm0.1$ & $6.4\pm0.3$  & $0.28$ & $2.6\pm0.1$ & $13.2\pm0.7$ & $0.19$ & $0.15$\\
R2 & $13\times13$ & -- & $3.7\pm0.2$  & -- & -- & $7.6\pm0.4$ & -- & -- \\ 
R3 & $7\times7$ & $0.9\pm0.05$ & $3.9\pm0.2$ & $0.23$ & $0.5\pm0.02$ & $10.1\pm0.5$ & $0.05$ & $0.77$\\
R4 & $7\times7$ & $0.7\pm0.04$ & $1.5\pm0.1$ & $0.47$ & $1.5\pm0.1$ & $3.4\pm0.2$ & $0.46$ & $0.03$\\
R5 & $13\times13$ & $0.5\pm0.03$ & $1.5\pm0.1$ & $0.35$ & $1.0\pm0.1$ & $3.2\pm0.2$ & $0.31$ & $0.12$ \\
I & $13\times13$ & -- & $1.6\pm0.2$  & -- & -- & $3.5\pm0.3$ & -- & --  \\ 
\hline
\end{tabular}
\end{center}
{Note: $^{\rm (a)}$ Uncertainties are of the same order of those on the total intensity which are given by $\sqrt{ (\zeta I_\lambda)^2 + \sigma_{{\rm rms,} I}^2 N_{\rm beam} }$ ($\zeta=0.05$ is the calibration uncertainty, $\sigma_{{\rm rms,} I}$ is the Stokes $I$ noise map and $N_{\rm beam}=A_{\rm source}/A_{\rm beam}$ is the number of beam in the source where we measure the flux).  $^{\rm (b)}$ Uncertainties are dominated by the precision on the leakage calibration (0.5\%, \url{https://science.nrao.edu/facilities/vla/docs/manuals/obsguide/modes/pol}).}
\label{tab:fluxes}
\end{table*}

We detect significant polarized emission both from the numerous radio galaxies and from the diffuse radio sources. The brightest polarized structure of the cluster is the northern relic (RN), with integrated polarized flux densities of $P_{\rm 3.0GHz}=17.0\pm0.9$ and $P_{\rm 1.5GHz}=19.4\pm1.0$ mJy (Table \ref{tab:fluxes}). The relic presents a similar continuous shape as detected in total intensity emission (see radio contours in Fig. \ref{fig:pol_emiss}). 
At $2.7''$ resolution (i.e. the highest resolution available in our observations), the polarized emission traces the relic's filamentary structure observed already in the total intensity \citep[see top panel in Fig. \ref{fig:RNzoom} in this manuscript and Fig. 7 in][]{digennaro+18}. 
Hints of polarized emission at $13''$ resolution are seen also from the very faint relic northward of RN, i.e. R5, with high degree of polarization at both 3.0 and 1.5 GHz (i.e. about 35\% and 30\%).

Particularly bright in polarization is also the relic located eastward of RN, i.e. R1 ($P_{\rm 3.0GHz}=1.5\pm0.1$ and $P_{\rm 1.5 GHz}=2.6\pm0.1$ mJy). The relic labelled as R4 shows a particularly high degree of polarization at both 3.0 and 1.5 GHz ($\sim50\%$), with negligible wavelength-dependent depolarization. On the contrary, the relic westward of RN, i.e. R3, undergoes strong depolarization from 3.0 to 1.5 GHz ($\rm DP^{3.0GHz}_{1.5GHz}\sim80\%$).

Faint polarized emission is observed in the southern relic (RS), at $13''$ resolution. Here, the emission only comes from two out of the five  ``arms'' that were detected in \cite{digennaro+18}, i.e. only RS1 and RS2. This is not completely a surprise, as these two ``arms'' are also the brightest in total intensity \citep[see][]{digennaro+18}. 

No polarized emission is detected for the diffuse sources R2 and I. Finally, we detect polarized emission from the radio galaxies in and around the cluster (i.e. A, B, C, D, E, F, H, J, K1, M, N and O), whose degree of polarization at 1.5 and 3.0 GHz ranges between 1--10\%, consistently with other similar objects \citep[e.g.][]{o'sullivan+12}.

\begin{figure*}
\centering
{\includegraphics[width=\textwidth]{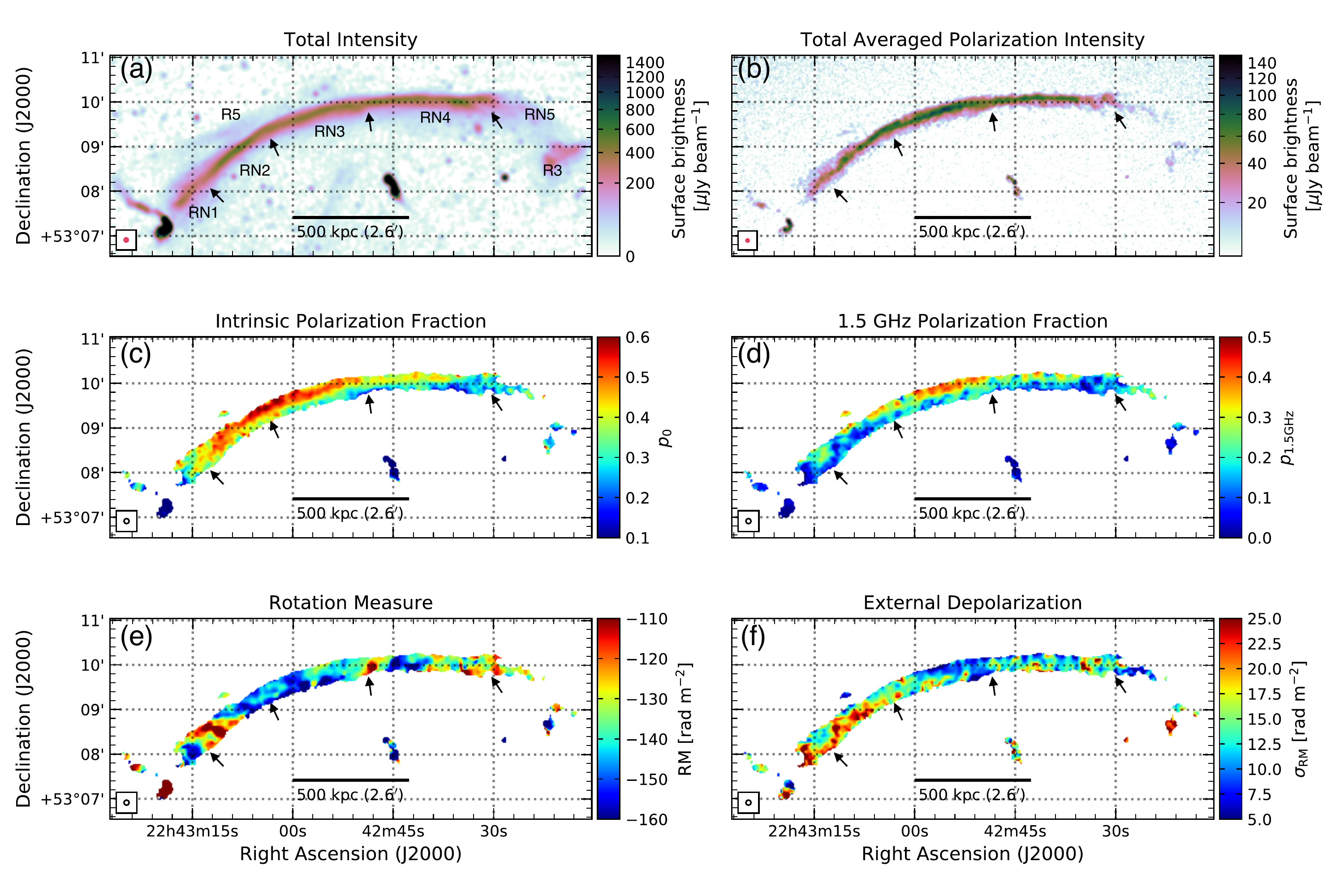}}
\caption{Panels (a) and (b): 1--4 GHz Stokes $I$ emission of the northern relic \citep{digennaro+18} and correspondent 1.26--3.60 GHz averaged polarized emission (not corrected for the Ricean bias) at $\sim5^{\prime\prime}$ resolution. Panels (c), (d), (e), and (f): 
intrinsic polarization fraction, polarization fraction at 1.5 GHz, Rotation Measure and External Depolarization maps at $7^{\prime\prime}$ resolution.
Black arrows in the plots are located at same physical coordinates, and indicate the points where the relic breaks into separate filaments \citep[see also Fig. \ref{fig:RNzoom} in this manuscript and Fig. 7 in][]{digennaro+18}. 
Uncertainty maps corresponding to panels (c) to (f) are displayed in Appendix \ref{apx:additonal}.}
\label{fig:RNpolarization}
\end{figure*}

\begin{figure*}
\centering
{\includegraphics[width=\textwidth]{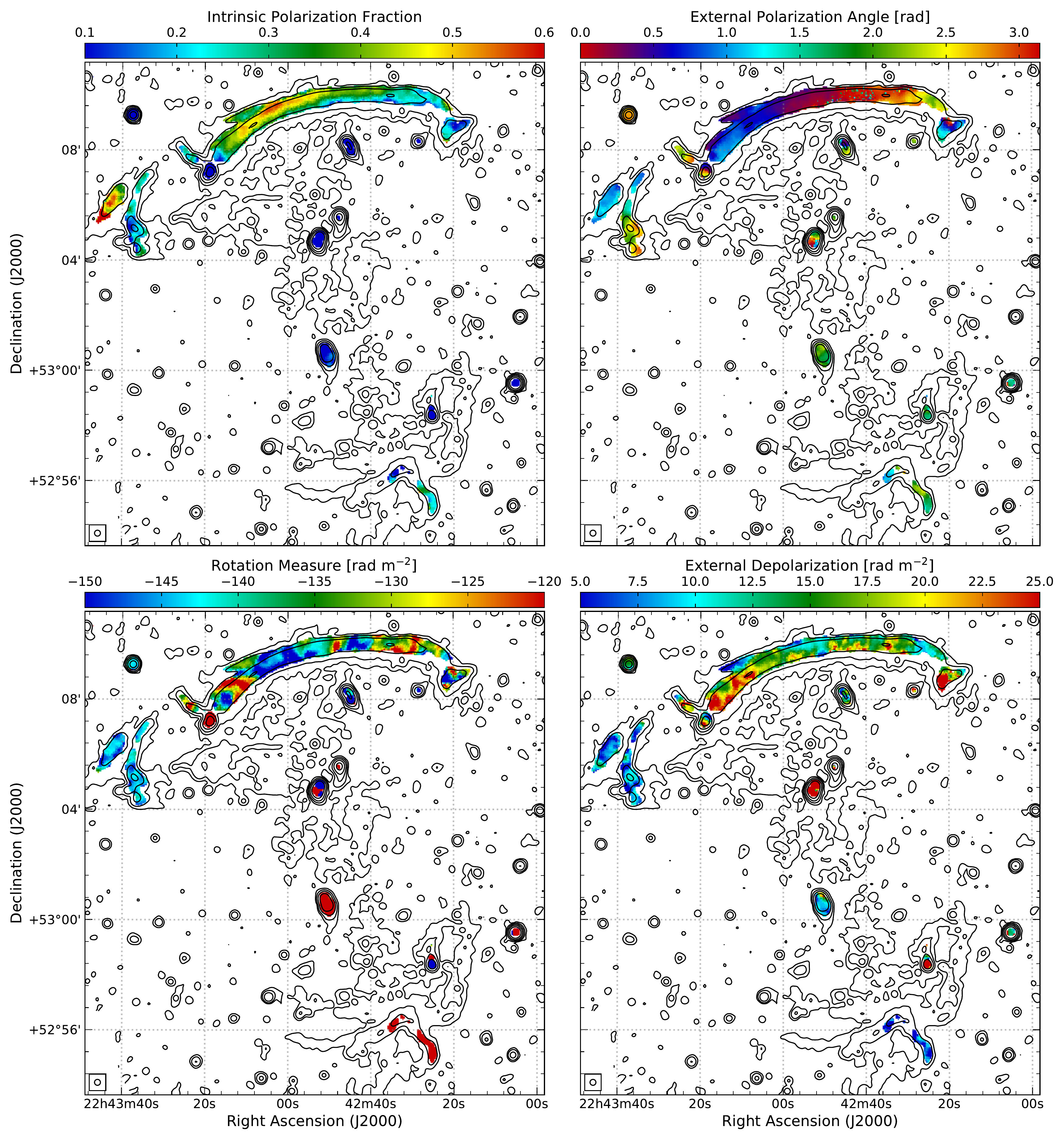}}
\caption{From top left to bottom right: intrinsic polarization fraction ($p_0$), intrinsic angle ($\chi_0$), Rotation Measure (RM) and depolarization ($\sigma_{\rm RM}$) maps of CIZAJ2242 at $13^{\prime\prime}$ resolution. Stokes I radio contours at the same resolution are drawn in black at levels of $3\sigma_{\rm rms}\times\sqrt{[1, 4, 16, 64, 256,\ldots]}$, with $\sigma_{\rm rms}=6.2~\mu$Jy~beam$^{-1}$ \citep{digennaro+18}. Negative and positive uncertainty maps are displayed in Appendix \ref{apx:additonal}.}
\label{fig:maps}
\end{figure*}

\subsection{Intrinsic fractional polarization, intrinsic polarization angle, RM and depolarization  maps}\label{sec:maps}

In Fig. \ref{fig:RNpolarization} we show a comparison between the total intensity and total averaged polarization maps of the northern relic at $7''$ resolution (panels (a) and (b), respectively), best-fit intrinsic and 1.5 GHz polarization fractions ($p_0$ and $p_{\rm 1.5GHz}$, panels (c) and (d) respectively), Rotation Measure (RM, panel (e)) and external wavelength-dependent depolarization ($\sigma_{\rm RM}$, panel (f)) maps.
The polarization best-fit parameter maps of the full cluster at $13''$ resolution is shown in Fig. \ref{fig:maps}. These result from the $QU$-fitting approach for the case of the External depolarization (Eq. \ref{eq:eq:ext_far}) for each pixel with averaged polarized emission above $f\times\sigma_{{\rm rms},P}$. Here, $\sigma_{{\rm rms},P}$ is obtained at the given resolution as the root mean squared level of the averaged polarized emission measured in a central, ``empty'' region of the cluster. We use $f=2$ for the $2.5^{\prime\prime}$-tapered images with \texttt{weighting=`uniform'} and $f=3$ for all the other resolutions and \texttt{weighting=`Briggs'}. The corresponding uncertainty maps are displayed in Appendix \ref{apx:additonal}.

The northern relic (RN) shows very high best-fit intrinsic polarization fraction values at the outermost edge, with the eastern side up to 60\% and the western side up to 40\% polarized. We also note a radial decreasing of $p_0$ towards the cluster center. The intrinsic polarization angles approximately follow the shock normal, which is assumed to be perpendicular to the Stokes $I$ edge, supporting the scenario where the magnetic field is aligned after the shock passage (see also bottom panel in Fig. \ref{fig:RNzoom}). The angles remain aligned also in the downstream region. The Rotation Measure value is not constant along the relic, it spans east to west from $\rm RM\sim-150$ rad m$^{-2}$ to $\rm RM\sim-130$ rad m$^{-2}$, respectively, with median value of about $-141$ rad m$^{-2}$. Given the large distance from the cluster center (i.e. $\sim 1.5$ Mpc), where the contribution of the ICM is likely low, we suggest that this median value is mostly associated with the Galactic foreground (see Sect. \ref{sec:galacticRM}). The variations in RM across the northern relic ($\sim30$ rad m$^{-2}$, have a dominant scale of $\sim15''-30''$, and we cannot distinguish, with the available data, whether this is due to fluctuations in our Galaxy or in the ICM (see Sect. \ref{sect:RMfluctuations}).
Similar east-west RM and $p_0$ variations were reported with Effelsberg observations at 4.85 and 8.35 GHz \citep{kierdorf+17}. To the contrary, the RM value measured in the western side of the relic ($\rm RM\sim-130~rad~m^{-2}$) differs from what has been found by the Sardina Radio Telescope at 6.6 GHz \citep[$\rm RM\sim-400~rad~m^{-2}$,][]{loi+17}.
No north-south best-fit intrinsic polarization gradient across the relic's width was found by either \cite{kierdorf+17} or \cite{loi+17}, although their observations suffer from much lower resolution (i.e., $90''$ and $2.9'$, respectively) which smoothed out any possible downstream gradient. 
Interestingly, we measure RM values of about $\rm -100~rad~m^{-2}$ where the relic breaks in the RN1-RN2 and RN3-RN4 filaments (see panel (e) in Fig. \ref{fig:RNpolarization}).
Finally, we do not find any particular east-west trend in the $\sigma_{\rm RM}$ behavior, with an overall value of $\sigma_{\rm RM}\sim15-20$ rad m$^{-2}$ (see panel (f) in Fig. \ref{fig:RNpolarization}).  
These values differ from the high-frequency observations, as \cite{kierdorf+17} did not measure any depolarization for the northern relic. 

The radio relic R4 is characterized by a very high best-fit intrinsic polarization fraction ($\sim55\%$), while it is lower for R1, R3 and R5 ($\sim20\%$). 
No clear gradients have been observed for these sources, except for R3 which shows hints of increasing values of $p_0$ towards the cluster center. The RM values are rather constant across R1 and R4, $\rm RM \sim -142$ rad m$^{-2}$, consistent with the one found for source N: since this radio galaxy is located outside of the cluster, its Rotation Measure is likely associated with the screen of our Galaxy rather than the ICM. Also, R1 and R4 have a very small values of $\sigma_{\rm RM}$, again consistent with their spatial position in the cluster, in a region of low ICM density.

In the southern relic (RS), we measure a relatively low best-fit intrinsic polarization fraction of $\sim10-25\%$. 
Across RS1 and RS2, the Rotation Measure spans from $\sim -90$ to $\sim -80$ rad m$^{-2}$. As for the northern relic, since RS is located in the cluster outskirts, we speculate that most of its RM is due to the Galaxy. The discrepancy between $\rm RM_{RN}$ and $\rm RM_{RS}$ can be either due to our Galaxy, whose RM variation is very uncertain (Sect. \ref{sec:galacticRM}), or to a different combination of $n_e B_\parallel$ along the line of sight northward and southward the cluster ICM (see Eq. \ref{eq:rm_fit}).

Finally, the polarized radio galaxies in the cluster field present different values of Rotation Measure. This possibly reflects the combination of their different position in the ICM with the Galactic contribution, although their intrinsic RM cannot be fully excluded. Among them, sources~D and~C are particularly interesting. They are located, in projection, in the cluster center and we measure a large difference in RM in the source's lobes, with the northwestern being negative (i.e. $\sim -600$ and $\sim -200$ rad m$^{-2}$, for source D and C respectively) and the southeastern being positive (i.e. $\sim +300$ and $\sim +250$ rad m$^{-2}$, for source D and C respectively). Such an extreme variation of RM in the lobes of the two radio galaxies probably originates in the radio galaxies themselves, although some effects might also be associated with the large amount of ICM traversed by the polarized emission. However, for these sources we find that a single-RM model does not properly fit the data, even within a single resolution element (i.e. a single pixel, see Appendix \ref{apx:dwnpix}). We therefore suggest the presence of a complex RM structure, as is observed also in other radio galaxies \citep[e.g.][]{o'sullivan+12}. This study is, however, beyond the scope of this paper.

\section{Discussion}\label{sec:disc}
Radio relics are thought to trace merger-induced shock waves which (re-)accelerate electrons and compress and amplify the cluster magnetic fields \citep[e.g.,][]{ensslin+98}. While several studies have been performed to investigate the mechanism to produce the highly-relativistic electrons in radio relics \citep[e.g.][]{brunetti+jones14,fujita+15,donnert+16,kang+17}, studies of their magnetic field properties have been challenging, mostly because depolarization effects are stronger at low frequencies (i.e. $\lesssim1$ GHz).

The northern radio relic in CIZAJ2242, i.e. the Sausage relic, is well-known to be highly polarized, hence it represents one of the best target for detailed polarization studies. 
Here, we present the first analysis of the radial and longitudinal polarization properties of the relic in the post-shock region on ten-kpc scales (i.e. $\sim8-40$ kpc). Additionally, we investigate possible correlations between the polarization parameters and look for the presence of possible underlying trends among them by calculating the running median along the $x$-axis, with moving boxes of 20 windows. The uncertainties are calculated as $\sigma_\pm/\sqrt{N}$, with $\sigma_+=y_{0.50}-y_{0.16}$ and $\sigma_-=y_{0.84}-y_{0.50}$ (with $y_{0.16}$, $y_{0.50}$ and $y_{0.84}$ the 16\%, 50\%, i.e. the median, and 84\% of the distribution, respectively), and $N$ the number of windows \citep{lamee+16}. The existence of a correlation was then evaluated by means of the Pearson coefficient, $r_p$ \citep[][]{pearson1895}, where we define $|r_p|\leq0.3$ as no/very weak correlation, $0.3 < |r_p|\leq0.7$ as weak/moderate correlation, and $|r_p|>0.7$ as strong correlation. We also report the Spearman coefficient, $r_s$, which assesses whether the relationship is monotonic (i.e. $|r_s|\leq 0.3$: no/very weakly monotonic ; $0.3 < |r_s| \leq 0.7$: weakly/moderately monotonic ; $|r_s|>0.7$: strongly monotonic).

The following discussion is focused on the Sausage relic. In Sect. \ref{sec:profiles} we present the radial profiles of the best-fit polarization parameters; in Sect. \ref{sec:p0profile} we discuss possible explanation for the profile found for the best-fit $p_0$; in Sect. \ref{sec:turb} we look at the contribution of the turbulent magnetic field in the post-shock region; in Sect. \ref{sec:depol} we investigate the limitation of the observing bandwidth coverage; finally, in Sect. \ref{sect:RMfluctuations} we look at the RM fluctuation in the relic.

\subsection{Polarization parameters radial profiles}\label{sec:profiles}
\begin{figure}
\centering
\includegraphics[width=0.48\textwidth]{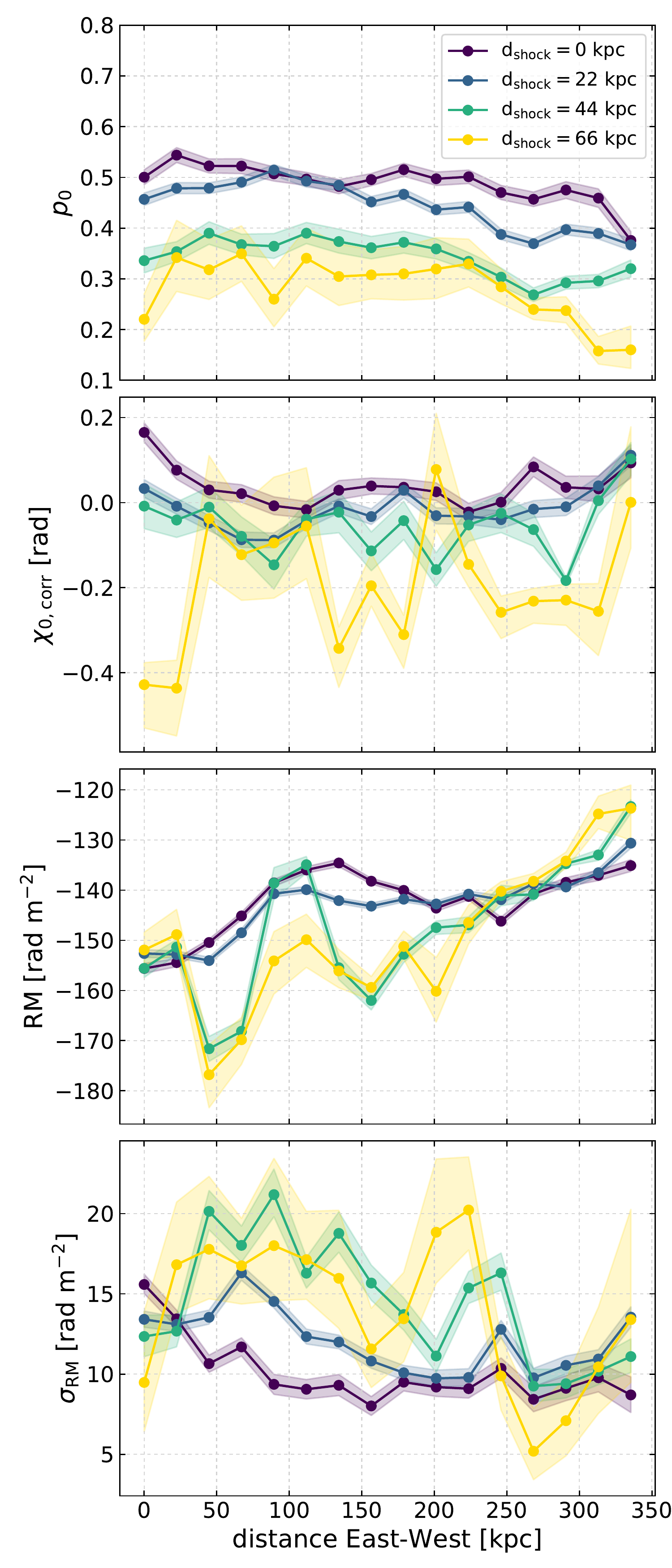}
\caption{From top to bottom: East-West profiles on the RN3 filament for the best-fit intrinsic polarization fraction ($p_0$), intrinsic polarization angle corrected for the shock normal ($\chi_{\rm 0, corr}$), Rotation Measure (RM) and depolarization ($\sigma_{\rm RM}$) using the External Faraday Rotation dispersion model (Eq. \ref{eq:eq:ext_far}). Different colors represent different distances from the shock ($d_{\rm shock}$, see legend), being the shock located at the  outermost edge of the relic, and the correspondent shaded areas show the uncertainties on the measurements.}\label{fig:profiles}
\end{figure}

\begin{figure}
\centering
\includegraphics[width=0.45\textwidth]{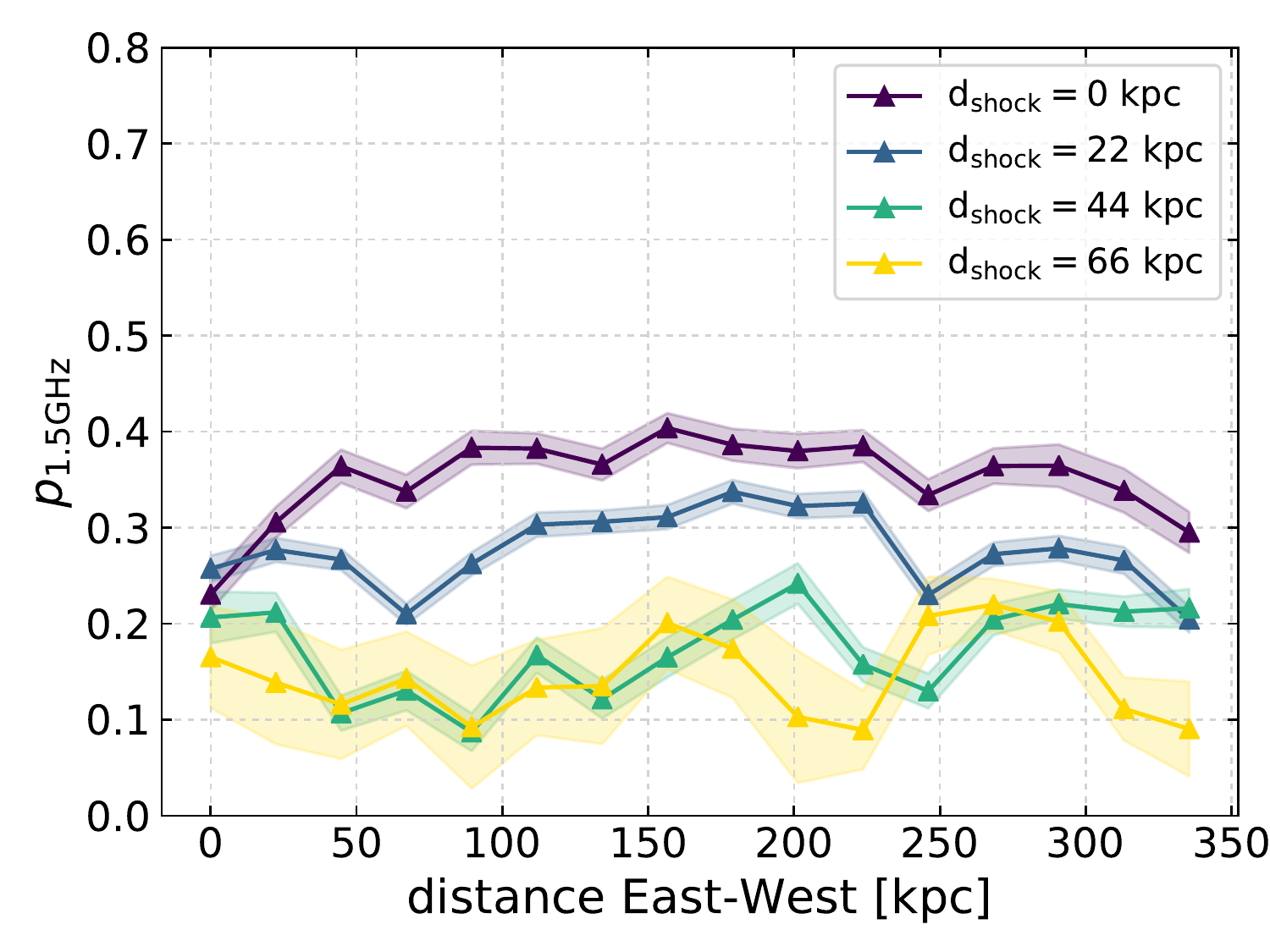}
\caption{As the top panel in Fig. \ref{fig:profiles}, but for the polarization fraction at 1.5 GHz.}
\label{fig:profiles_1.5}
\end{figure}

We repeated the $QU$ fit using Eq. \ref{eq:eq:ext_far} in beam-sized boxes (i.e. $7''$, resulting in a linear size of about 20~kpc at the cluster redshift, see legend in Figs. \ref{fig:profiles} and \ref{fig:profiles_1.5}, and Fig. \ref{fig:boxesRN3}) covering the filament RN3, which we consider to be representative part of the relic (see Fig. \ref{fig:RNpolarization}). For each single radial annulus (i.e. same-colored markers in Figs. \ref{fig:profiles} and \ref{fig:profiles_1.5}), the polarization parameters have a similar trend along the filament (i.e. east to west, Fig. \ref{fig:profiles}), with the exception for the Rotation Measure which shows a variation of about 30 rad m$^{-2}$. On the other hand, a clear north-south trend is visible for the best-fit intrinsic polarization fraction. It drops about 35--40\%, from an average value of $\langle p_0 \rangle_{d=0{\rm kpc}} = 0.40\pm0.04$ at the shock position to $\langle p_0 \rangle_{d=66{\rm kpc}} = 0.28\pm0.06$ in the innermost downstream annulus (top panel in Fig. \ref{fig:profiles}). The same trend is also observed for the polarization fraction at 1.5 GHz (Fig. \ref{fig:profiles_1.5}). At this wavelength, the drop is even larger,  about 60\% (from $\langle p_{\rm 1.5 GHz} \rangle_{d=0{\rm kpc}} = 0.35\pm0.04$ to $\langle p_{\rm 1.5 GHz} \rangle_{d=66{\rm kpc}} = 0.24\pm0.09$). A similar but opposite trend is observed for the external wavelength-dependent depolarization: here we found higher values towards the downstream region (from $\langle \sigma_{\rm RM} \rangle_{d=0{\rm kpc}} = 10.1\pm0.2$ to $\langle \sigma_{\rm RM} \rangle_{d=66{\rm kpc}} = 13.9\pm0.8$ rad m$^{-2}$, bottom panel in Fig. \ref{fig:profiles}). Hints of these radial trends are also seen in the entire relic (Fig. \ref{fig:p0_spix}; see Appendix \ref{apx:grid} for a view on the beam-sized boxes where we performed the $QU$ fit). In this case, the radial information is obtained by looking at the spectral index, $\alpha^{\rm 150MHz}_{\rm 3.0GHz}$, since steeper values are located further in the downstream region where synchrotron and Inverse Compton energy losses increase \citep[e.g.,][]{digennaro+18}. We calculated $\alpha^{\rm 150MHz}_{\rm 3.0GHz}$ using the LOFAR (150 MHz), GMRT (610 MHz) and VLA (1.5 and 3.0 GHz) maps described in \cite{hoang+17}, \cite{vanweeren+10} and \cite{digennaro+18}, respectively.  We found Pearson and Spearman rank coefficients of $r_p=-0.28$ and $r_s=-0.28$ for the $p_0$--$\alpha^{\rm 150MHz}_{\rm 3.0GHz}$ distribution, and $r_p=0.16$ and $r_s=0.24$ for the $\sigma_{\rm RM}$--$\alpha^{\rm 150MHz}_{\rm 3.0GHz}$ distribution. These measurements show, for the first time, that the northern relic in CIZAJ2242 suffers from both wavelength- and radial-dependent depolarization.

Finally, no clear downstream variations are seen for the intrinsic polarization angle corrected for the shock normal in the plane of the sky\footnote{Uncertainties on $\chi_{\rm 0, corr}$ are determined included the uncertainties on $\chi_0$ ($\sim0.01$ rad, from the fitting procedure using MCMC) and on $n$ within the beam region ($\sim0.02$ rad at $7''$ resolution).} ($\chi_{\rm 0, corr}=\chi_0-n$, second panel in Fig. \ref{fig:profiles}) and for the Rotation Measure (third panel in Fig. \ref{fig:profiles}; see also Sect. \ref{sect:RMfluctuations}).

\begin{figure*}
\centering
{\includegraphics[width=0.45\textwidth]{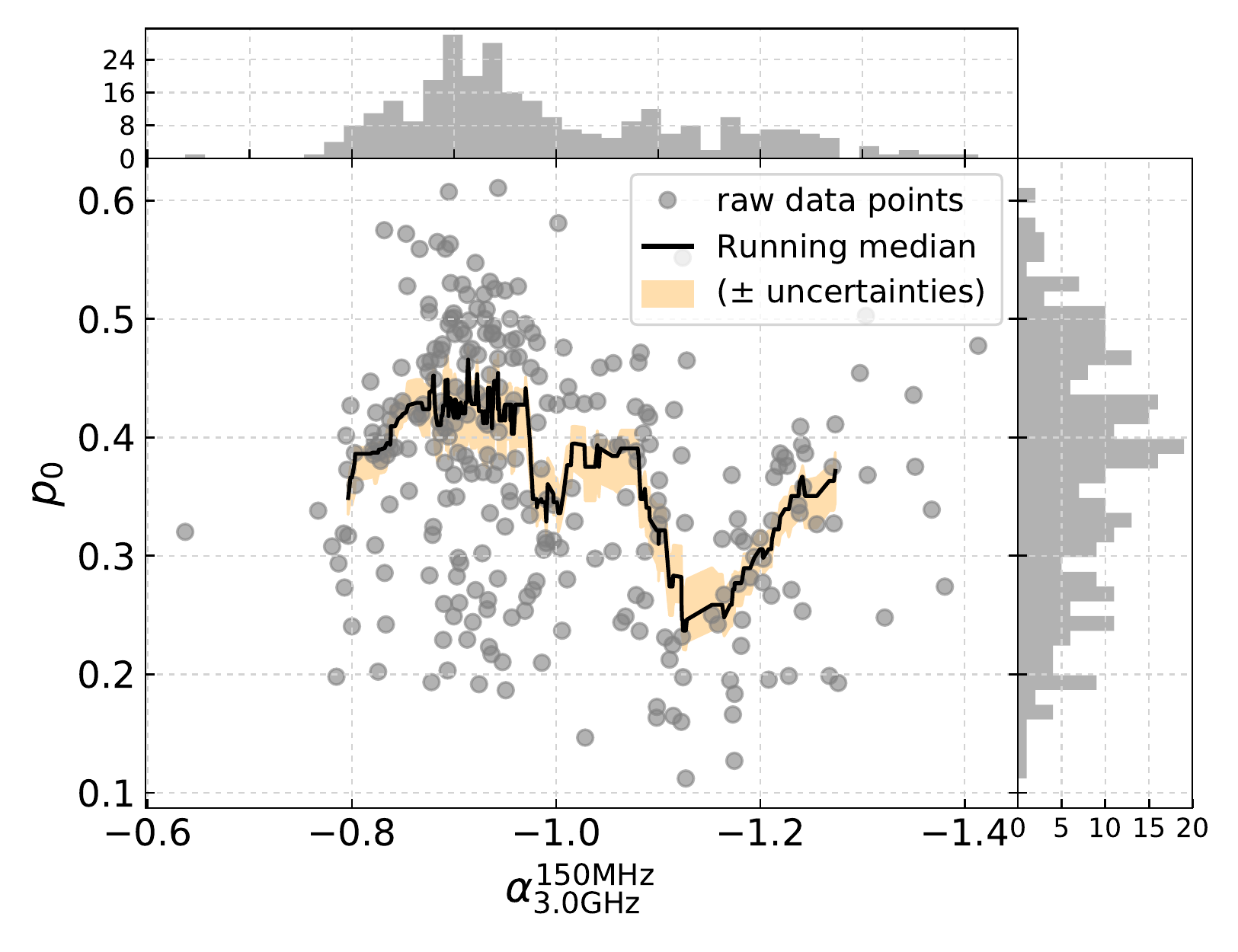}}
{\includegraphics[width=0.45\textwidth]{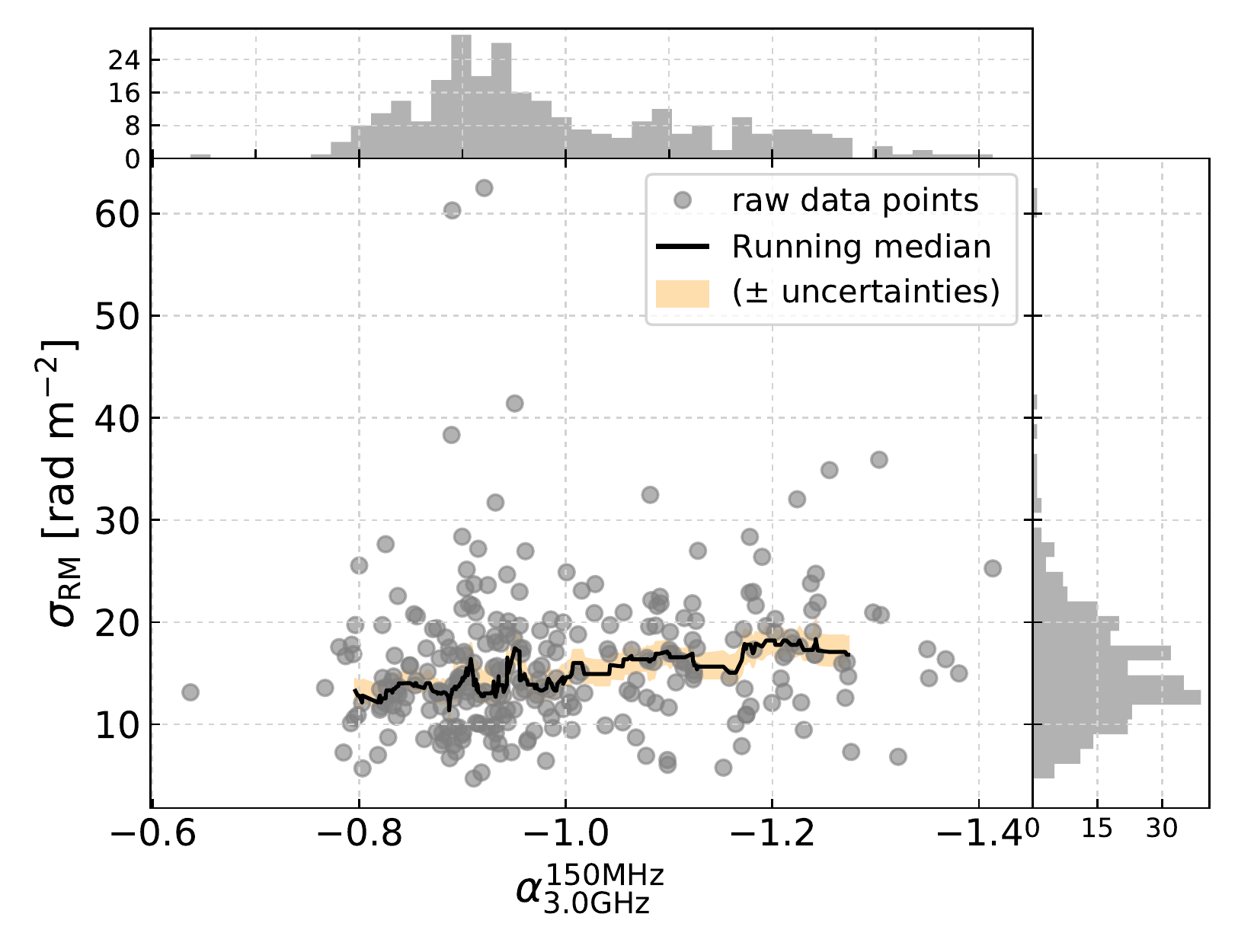}}
\caption{Distributions of the intrinsic polarization fraction and external wavelength-dependent depolarization as a function of the spectral index (grey circles in the  left and right panel, respectively). The grey histograms show the projected distribution of the $y$- and $x$-axis quantities along each axis. The black solid line shows the running median of $p_0$ and $\sigma_{\rm RM}$ calculated using 20 windows in the $\alpha^{\rm 150MHz}_{\rm 3.0 GHz}$ space, while the yellow area represents the correspondent uncertainties.}\label{fig:p0_spix}
\end{figure*}

\subsection{On the downstream depolarization}\label{sec:p0profile}
In the following sections, we discuss two possible explanations for the observed radial profile of the polarization fraction. In particular, we investigate the role of wavelength-dependent depolarization and Faraday Rotation (Sect. \ref{sec:FR_depol}) and include a three-dimensional modelling of the relic (Sect. \ref{sec:geom}).

\subsubsection{Wavelength-dependent depolarization and Faraday Rotation effects}\label{sec:FR_depol}
A naive explanation for the downstream depolarization is the effect of a complex magneto-ionic layer that might differently rotate  the polarization vectors in different parts of the relic. According to this scenario, the bottom panel in Fig. \ref{fig:profiles} and the right panel in Fig. \ref{fig:p0_spix} both suggest a mild increasing contribution of the external wavelength-dependent depolarization in the downstream region.

\begin{figure*}
\centering
{\includegraphics[width=0.45\textwidth]{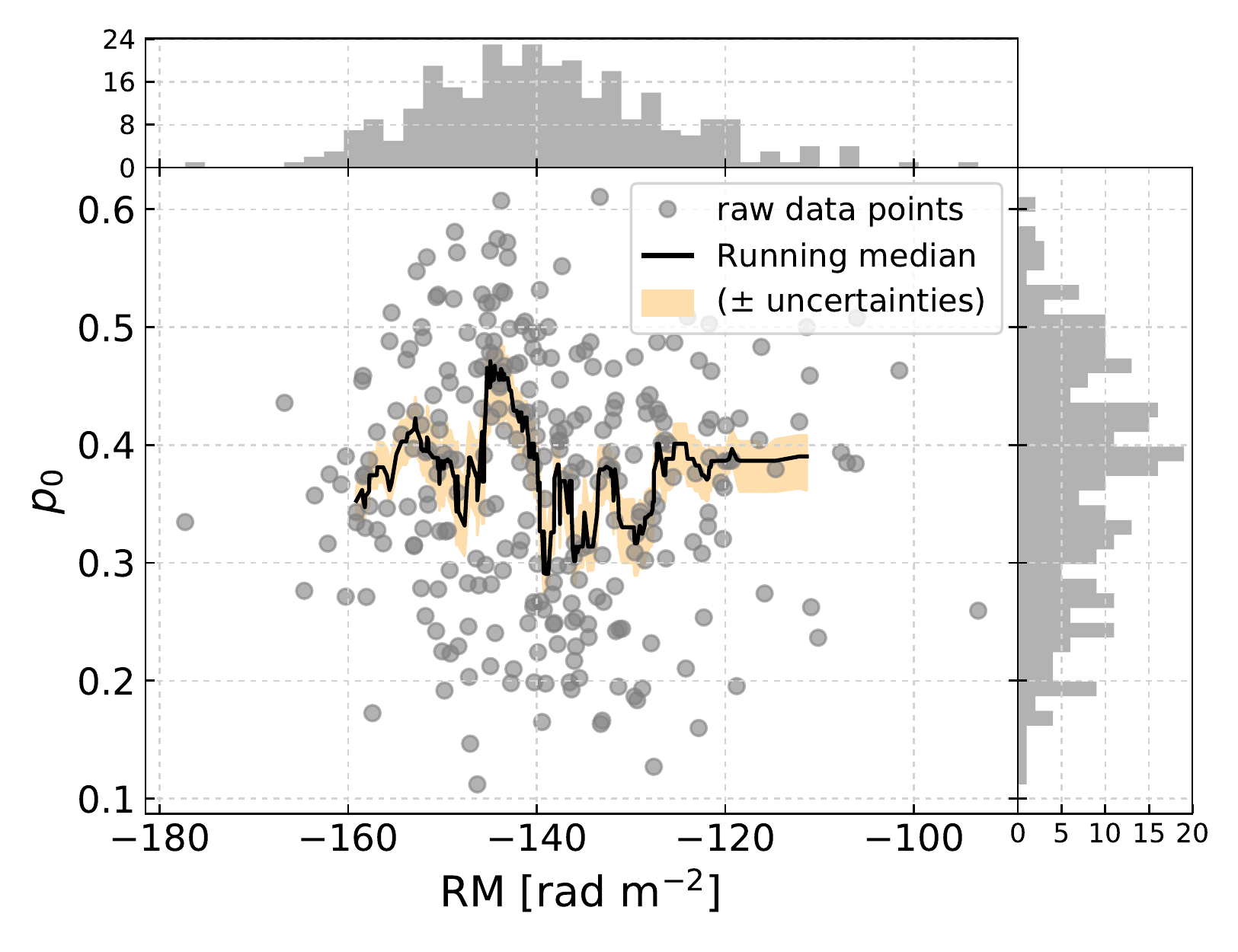}}
{\includegraphics[width=0.45\textwidth]{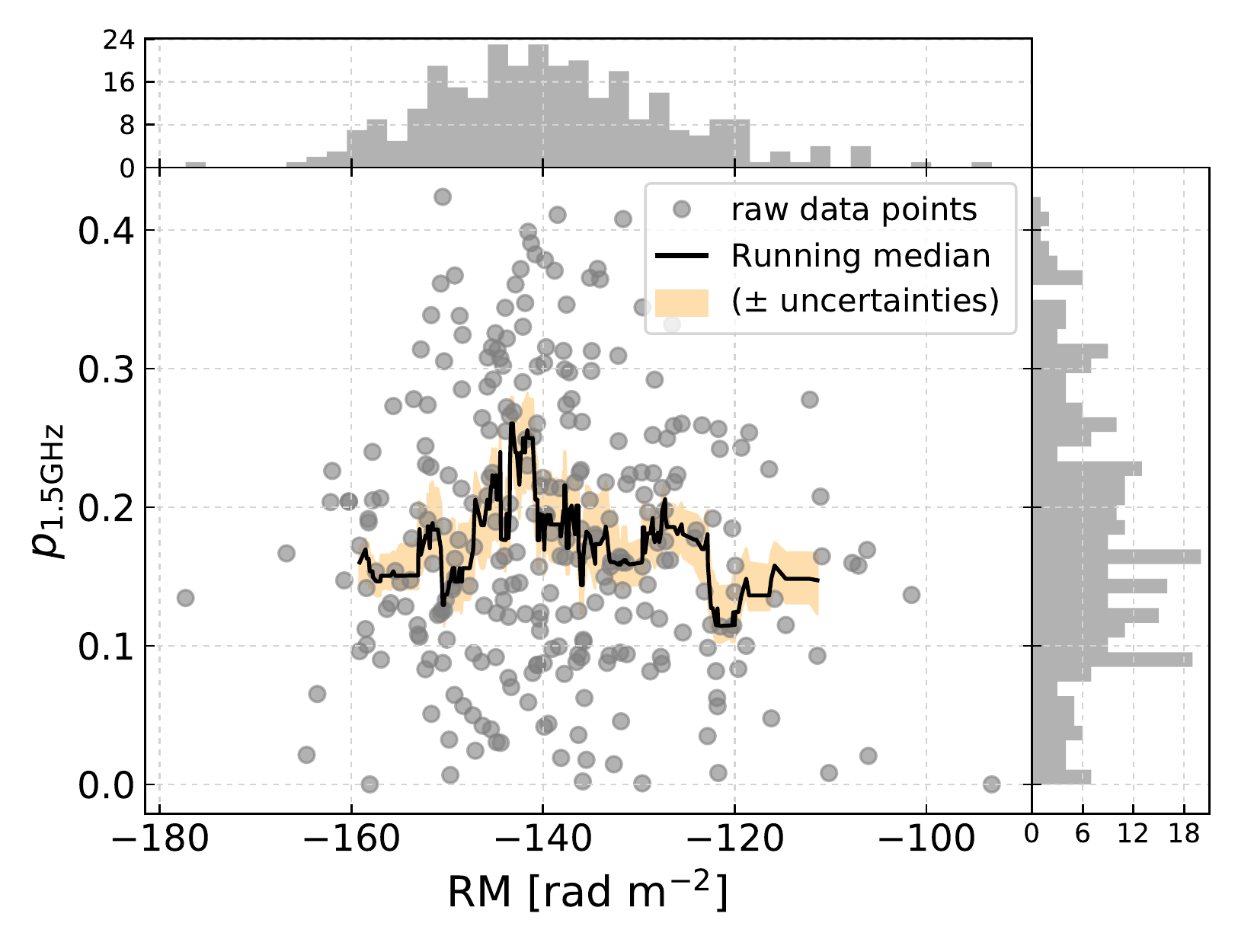}}\\
{\includegraphics[width=0.45\textwidth]{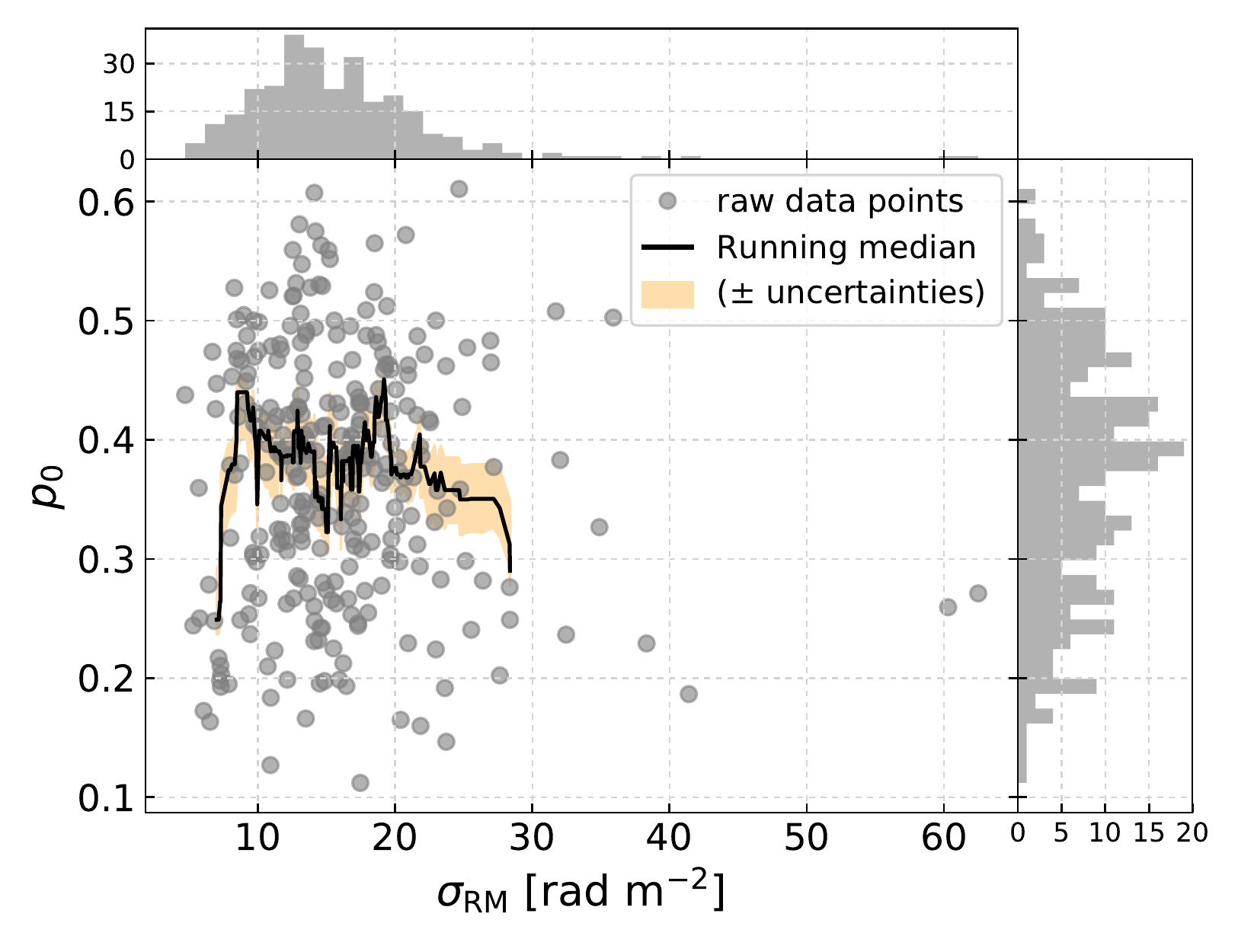}}
{\includegraphics[width=0.45\textwidth]{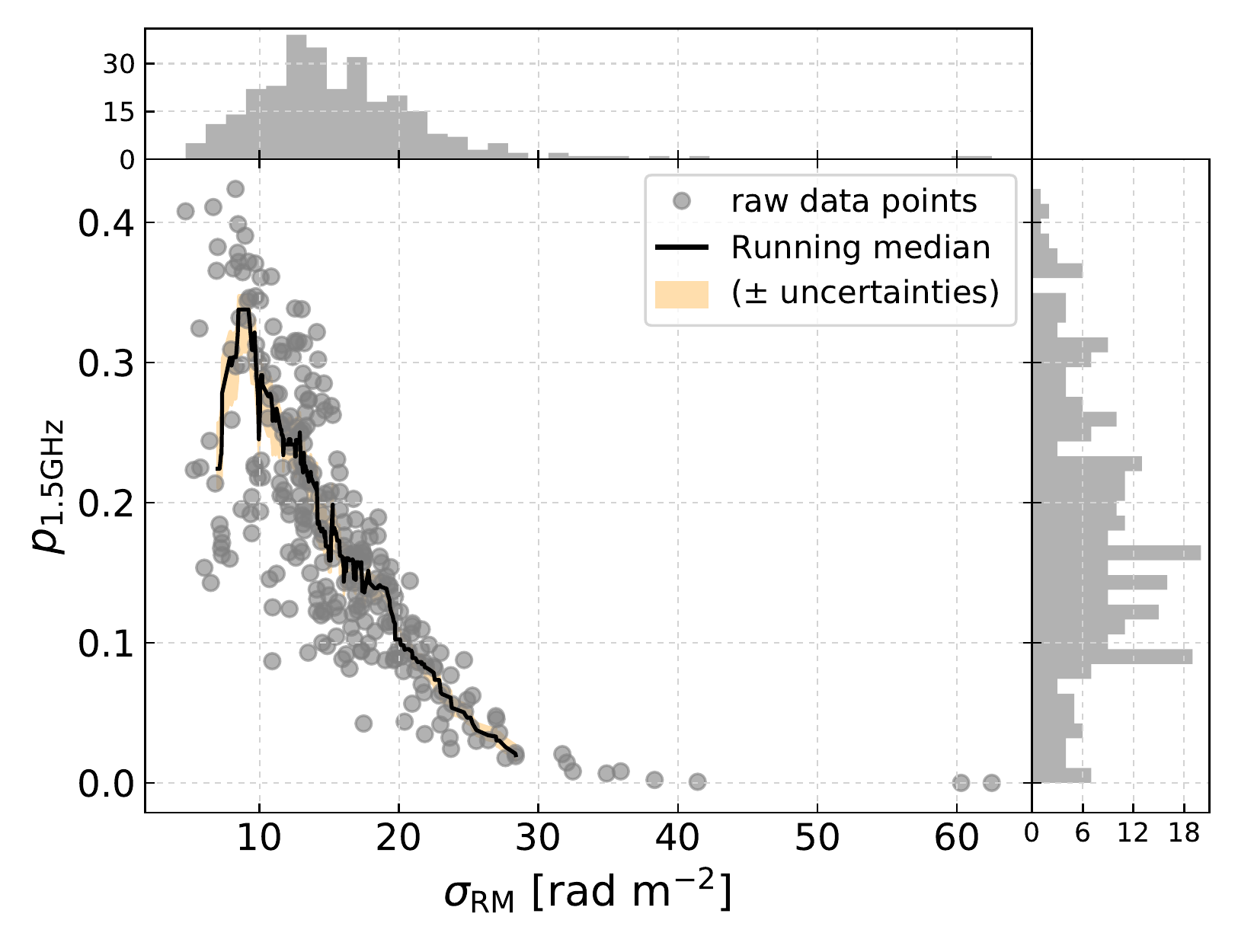}}
\caption{Distributions of the intrinsic and 1.5 GHz  polarization fractions (left and right column respectively) as a function of the absolute relative Rotation Measure and external wavelength-dependent depolarization (grey circles in the top and bottom panels, respectively). The grey histograms show the projected distribution of the $y$- and $x$-axis quantities along each axis. For both columns, the solid black line represents the running median of the $y$-axis variable (i.e. $p_0$ and $p_{\rm 1.5 GHz}$) calculated using 20 windows in the space of the $x$-axis variable (i.e. RM and $\sigma_{\rm RM}$). The yellow shaded area represents the uncertainty on the running median.}
\label{fig:corr_p0_p1.5}
\end{figure*}

\begin{table}
\caption{Pearson ($r_p$) and Spearman ($r_s$) rank correlation coefficients of the running median in Figs. \ref{fig:p0_spix} and \ref{fig:corr_p0_p1.5}.}
\begin{center}
\begin{tabular}{lcc}
\hline
\hline
Parameters & $r_p$ & $r_s$ \\
\hline	
$p_0$--RM & $-0.06$ & $-0.09$ \\
$p_0$--$\sigma_{\rm RM}$ & $-0.06$ & $-0.01$\\
$p_{\rm 1.5GHz}$--RM & $-0.07$  & $-0.04$\\
$p_{\rm 1.5GHz}$--$\sigma_{\rm RM}$ & $-0.73$ & $-0.82$ \\
$p_0$--$\alpha^{\rm 150MHz}_{\rm 3.0GHz}$ & $-0.28$ & $-0.28$ \\
$\sigma_{\rm RM}$--$\alpha^{\rm 150MHz}_{\rm 3.0GHz}$ & $0.16$ & $0.24$ \\
\hline
\end{tabular}
\end{center}
\label{tab:spearm_coeff_p0_p1.5}
\end{table}

We investigated the relation between the best-fit intrinsic polarization fraction and the measured Rotation Measure and external wavelength-dependent depolarization (left column in Fig. \ref{fig:corr_p0_p1.5}). In both cases, we do not see particular trends, nor underlying fluctuations from the analysis of the running median. Both the Pearson and Spearman rank coefficients confirm the visual inspection, being $r_p=-0.06$ and $r_s=-0.09$ for the $p_0$--RM distribution and $r_p=-0.06$ and $r_s=-0.01$ for the $p_0$--$\sigma_{\rm RM}$ one (see Table~\ref{tab:spearm_coeff_p0_p1.5}). We therefore conclude that our best-fit intrinsic polarization fraction is independent from external factors, as the Faraday Rotation and the wavelength-dependent depolarization. On the other hand, an anti-correlation in the $p_{\rm 1.5 GHz}$--$\sigma_{\rm RM}$ distribution is observed ($r_p=-0.73$ and $r_s=-0.82$). No correlation has been found for the $p_{\rm 1.5GHz}$--RM one ($r_p=-0.07$ and $r_s=-0.04$). These suggest that only the wavelength-dependent depolarization affects the polarization fraction at lower frequencies.

\subsubsection{Relic three-dimensional shape}\label{sec:geom}
\begin{figure}
\centering
\includegraphics[width=0.45\textwidth]{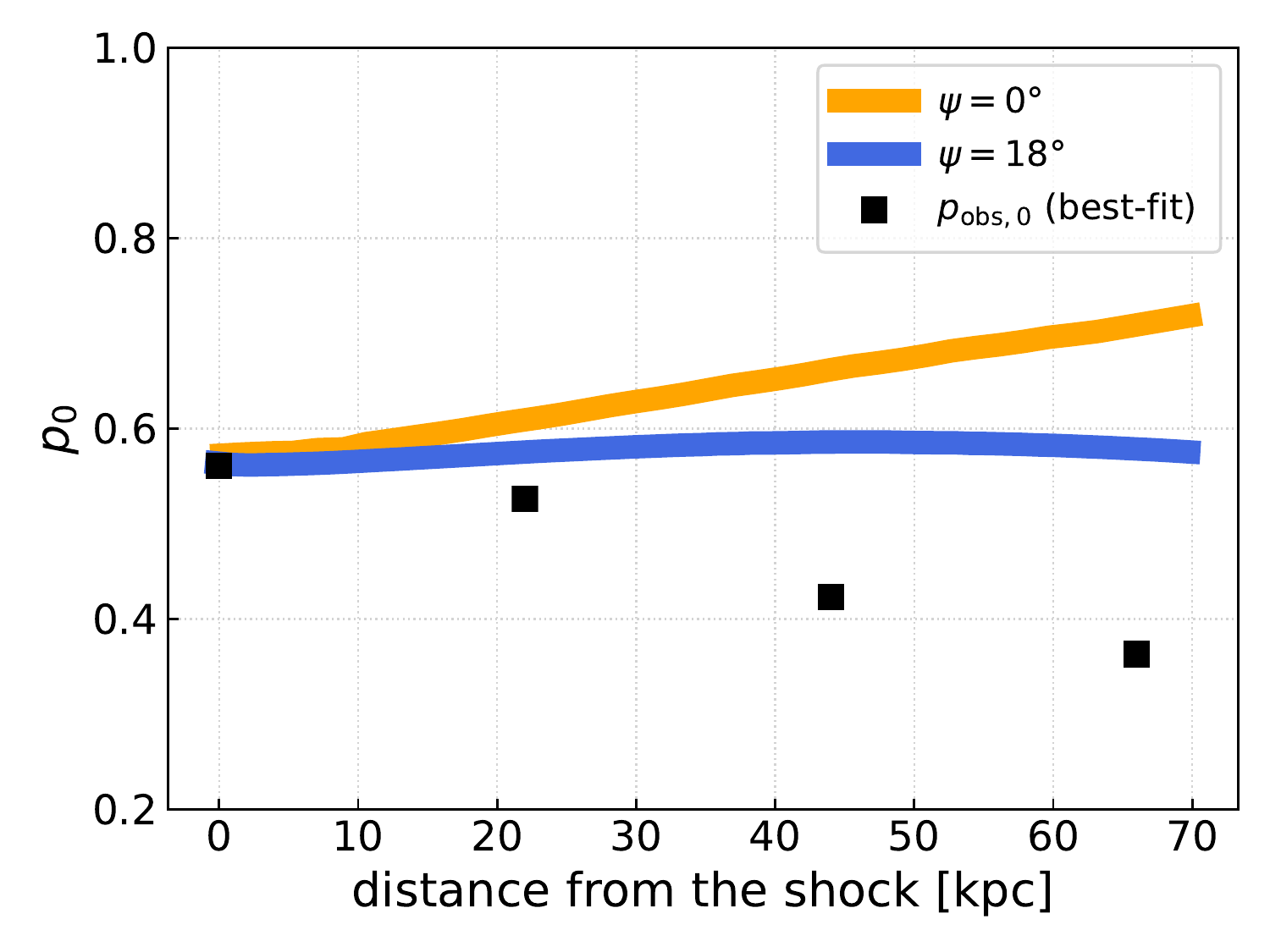}
\caption{Theoretical profiles of the intrinsic polarization fraction in the post-shock region assuming a shock wave perfectly aligned with the plane of the sky (i.e. $\psi=0^\circ$, orange line) and assuming an opening angle for the relic of $\psi=18^\circ$ \citep[blue line]{digennaro+18}.
Black squares represent the best-fit intrinsic polarization fraction values obtained from a smaller sector of RN3 (i.e. where we could assume constant polarization parameters in the east-west direction).
}
\label{fig:geomproj}
\end{figure}

For a power law electron energy distribution with slope  $\delta=1-2\alpha$, i.e. $dN(E)/dE \propto E^{-\delta}$, in a region with homogeneous magnetic field the intrinsic polarisation amounts to \citep{rybicki+ligthman86}:
\begin{equation}\label{eq:intpol_spix}
p_0 = \frac{3 \delta + 3 }{3 \delta + 7} \, .
\end{equation}
Therefore, if the slope of the electron distribution varies across the relic the intrinsic polarisation will also vary. According to the standard scenario for relic formation, electrons are (re-)accelerated at the shock front, with a power law energy distribution, and cool subsequently due to synchrotron and Inverse Compton energy losses. Locally, the resulting electron spectrum may show a break, even if the sum of all these spectra is a power law again, \citep[see][for a detailed spectral analysis of the relic]{digennaro+18}. The locally curved spectra thus show a different intrinsic degree of polarization than the overall relic. From Eq. \ref{eq:intpol_spix}, the downstream region with the aged electron population would have a higher intrinsic polarisation fraction (orange line in Fig. \ref{fig:geomproj}).

Although the decreasing radial profile of the best-fit polarization degree seems to be in contrast with the above description, the complex shape of the shock front and the downstream region may impact the polarization, for instance by an inhomogeneous intrinsic polarisation fractions and by large differences in the path through the magnetized ICM from the emission to the observer. In this context, to reproduce a correct projected intrinsic polarization profile, it is necessary to take into account a realistic shape of the shock front, which has to include the contribution of its inclination with respect to the line of sight (M. Hoeft et al. in prep.). 

Following \cite{digennaro+18}, we created a toy model assuming that the shock front is a spherically symmetric cap in the plane determined by the line of sight and the cluster center, with a curvature radius of 1.5 Mpc and opening angle of $2\psi=36^\circ$ \citep[see also Fig. 10 in][]{kierdorf+17}. 
The alignment of electric field vectors with the shock normal (bottom panel in Fig. \ref{fig:RNzoom}) implies that the magnetic field is dominantly tangled on scales smaller than the resolution of the observations (i.e. $2.7''$). If the polarization angle reflects the structure of the magnetic field, we can assume a shock-compression scenario to explain the polarization properties of the relic \citep{ensslin+98}. In this scenario, an upstream isotropically tangled magnetic field is compressed by the shock front resulting in a downstream anisotropically tangled field, causing polarized synchrotron emission. In the specific case of RN, we adopt a shock Mach number of 3.7 which corresponds to an intrinsic polarization fraction of 58\%, when the shock is observed perfectly edge on. This value matches the maximum $p_0$ we estimated in the relic (see panel (c) in Fig. \ref{fig:RNpolarization}). The emission of different parts of the shock front is summed up, taking into account the angle between the shock normal and the line of sight, $90^\circ-\psi$. The more this angle deviates from $90^\circ$ the lower the intrinsic polarization becomes. Since those parts of the shock which deviate more from $90^\circ$ are shifted further downstream with respect to the outermost edge of the relic, the intrinsic polarization fraction decreases towards the downstream. For our model parameters, these two effects, namely the downstream increase in polarization due to the aging of the electrons population and the decrease due to the shift of those parts of the shock which are not seen perfectly edge on, cancel out, resulting in an almost constant theoretical $p_0$ profile. This, however, still deviates from our observations (see blue line and black squares in Fig. \ref{fig:geomproj}).

It is worth noting that we have used here a very simplified geometrical model that, for instance, does not explain the east-west $p_0$ variation we observed in the relic. 
Moreover, it does not include the effect of emitting regions at different Faraday depths in the relic downstream. According to the spherical model described above, at a distance of 60 kpc of the outer edge, the emission from the ``back side'' of the cluster travels about 800 kpc through the magnetised ICM, which causes additional downstream depolarization.  Interestingly, no evidence of multiple-RM components in the downstream region are observed in our data (see Appendix \ref{apx:dwnpix}). This suggests either that the relic cannot be described simply by a smooth spherical cap (e.g. overlapping filamentary structures) or we might be actually observing only the front/back side of the radio relic. On the other hand, the geometrical projections involve a number of adjustable parameters \citep[see, e.g.][]{kang+12}. Hence, a detailed modeling, which should include the shock shape, its downstream spectral and polarized characteristics and its physical properties \citep[such as the Mach number distribution, e.g.][]{ha+18,botteon+20}, is complicated and needs to be further examined.

\subsection{Turbulent magnetic field in the post-shock region}\label{sec:turb}
In the presence of both ordered and random magnetic field, Eq. \ref{eq:intpol_spix} can be written as \citep{sokoloff+98,govoni+04}:
\begin{equation}\label{eq:intpol_B}
p_0 = \dfrac{3\delta+3}{3\delta+7}\dfrac{1}{ 1+ \left ( \dfrac{B_{\rm rand}}{B_{\rm ord}} \right )^2}\, ,
\end{equation}
where $B_{\rm ord}$ represents the magnetic field component that is aligned with the shock surface and $B_{\rm rand}$ represents the isotropic magnetic field component. Thus, the ratio $B_{\rm rand}/B_{\rm ord}$ describes the order of isotropy of the magnetic field distribution.

In the northern relic of CIZAJ2242, the polarization angle seems to follow well the shock normal (see bottom panel in Fig. \ref{fig:RNzoom}), and no change is observed in the downstream region (second panel in Fig. \ref{fig:profiles}). This suggests that the component of the magnetic field parallel to the polarization angle is approximately constant in the downstream region. However, our measurements are limited by the observing resolution, which can hide the presence of tangled magnetic field on smaller scales and lead to a decreasing polarization fraction. If this is the case, from Eq. \ref{eq:intpol_B}, we can relate the radial decrease of $p_0$ with the decrease of the degree of anisotropy in the downstream region (i.e. the ratio $B_{\rm rand}/B_{\rm ord}$ increases). 
Given the averaged values found in the RN3 filament, i.e. $\langle p_0 \rangle_{d={\rm 0kpc}}\sim0.49$ and $\langle p_0 \rangle_{d={\rm 66kpc}}\sim0.28$, and assuming $\delta=3$ (i.e. $\alpha=-1$) we find that the ratio $B_{\rm rand}/B_{\rm ord}$ should increase of about 40\% in the downstream region.
Shock propagation in the ICM generates vorticity which boosts turbulence and amplify the magnetic field \citep[e.g.,][]{ryu+08}. Behind the shock, turbulence behaves more or less as a ``decaying'' turbulence, \citep[see, e.g.,][]{porter+15,donnert+18}, which might lead to the decreasing degree of anisotropy. Further studies are needed, however, upon this point.

\begin{figure}
\centering
\includegraphics[width=0.48\textwidth]{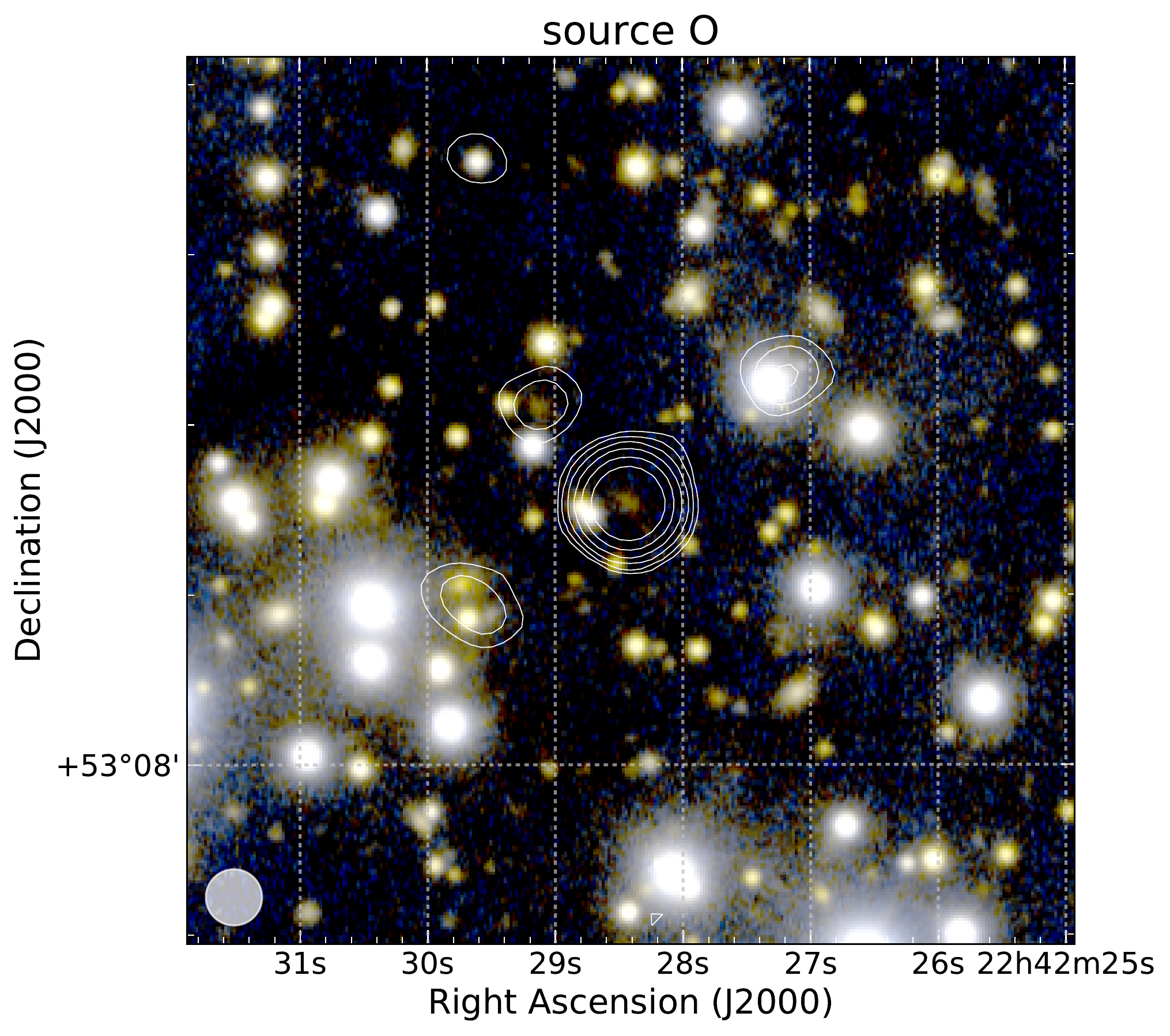}
\caption{Subaru $g$-$gi$-$i$ optical image of source O \citep{dawson+15,jee+15}. 1--4 GHz total intensity radio contours at $2.5''$ resolution are overlaid at levels of $3\sigma_{\rm rms}=\sqrt{(1,~4,~16,\dots)}$, with $\sigma_{\rm rms}=5.6~\mu$Jy beam$^{-1}$ the map noise \citep{digennaro+18}.}
\label{fig:opt_sourceO}
\end{figure}

The turbulent magnetic field $B_{\rm turb}$ is related to the wavelength-dependent depolarization, according to \citep{sokoloff+98,kierdorf+17}:

\begin{equation}\label{eq:Brand}
\sigma_{\rm RM} = 0.81 \sqrt{\frac{1}{3}} \langle n_e \rangle B_{\rm turb} \sqrt{\frac{L\Lambda}{f}}\, ,
\end{equation}
where $\langle n_e \rangle$ is the average electron density in cm$^{-3}$, $f$ is the volume filling factor of the Faraday-rotating gas, $L$ is the path length through the thermal gas and $\Lambda$ is the turbulence scale, both in pc unit. In the cluster area, only source O is a background polarized radio galaxy (see Fig. \ref{fig:opt_sourceO}). From our $QU$ fit, we found that the amount of the external depolarization for this source is very similar to that in RN, i.e. $\sigma_{\rm RM}\sim22$ rad m$^{-2}$ (see panel (f) in Fig. \ref{fig:RNpolarization} and bottom left panel in Fig. \ref{fig:maps}). Given the proximity of source O and RN and assuming that there is no contribution to the depolarization from source O itself and from the Galactic plane, we can use this $\sigma_{\rm RM}$ in Eq. \ref{eq:Brand} to obtain an approximate estimation of the tangled magnetic field in the northern relic, being $B_{\rm turb}\sim5.6~\mu$Gauss.  
Here, we used $\langle n_e \rangle=10^{-4}$ cm$^{-3}$ \citep{ogrean+14}, $L=350$ kpc\footnote{The path length of the magnetized plasma crossed by the polarized emission is $L\approx2\sqrt{2d_s\,r_s}$, where $d_s=10$ kpc and $r_s=1.5$ Mpc are the intrinsic width of the shock and its distance from the cluster center, respectively \citep[see][]{kierdorf+17}.}, $f=0.5$ \citep{govoni+04,murgia+04} and $\Lambda=8$ kpc\footnote{This is about one order of magnitude smaller than what is commonly used for galaxy clusters \citep[i.e. 100 kpc, see][]{iapichino+bruggen12}.}, i.e. the linear scale of our best resolution observation (i.e. $2.7''$). Note that the estimated $B_{\rm turb}$ is consistent with the upper value of the total magnetic field strength quoted by \cite{vanweeren+10}, leading to a ratio of magnetic and the thermal pressures $P_{\rm mag}/P_{\rm th}\sim0.11$ \citep{akamatsu+15}.

\subsection{Effect of the limited frequency-band coverage}\label{sec:depol}
The basic assumption of the $QU$-fitting approach is that, given observations in a wide band $\Delta\lambda^2=\lambda_{\rm max}^2-\lambda_{\rm min}^2$ and assuming a theoretical model, one can extrapolate the intrinsic polarization parameters, $p_0$ and $\chi_0$, at the ideal wavelength $\lambda\rightarrow 0$ where no wavelength-dependent effects (e.g. depolarization or Faraday Rotation) occur. The wider $\Delta\lambda^2$ and lower $\lambda_{\rm min}^2$ the better one can validate the theoretical model.  
However, due to the lack of high-resolution information at higher frequencies we cannot exclude the possibility of the existence of a more complex model to describe the polarized emission in RN. For example, \cite{ozawa+15} found a step-like fractional polarization profile in the radio relic in Abell 2256, with the fractional polarization increase occurring above 3.0 GHz. However, it is important to note that the presence of more complex models would result in a strong deviation from the Burn model in the downstream region, where a larger amount of magnetized plasma (i.e. the ICM) is crossed. Despite the low S/N, however, we see that the Burn approximation still holds in this region. 
Finally, $\Delta\lambda^2$ also sets the amount of wavelength-dependent depolarization detectable. Given our observing band, it would be rather difficult to determine $p(\lambda^2)$ if $\sigma_{\rm RM}\geq 100$ rad m$^{-2}$.

\begin{figure}
\centering
{\includegraphics[width=0.45\textwidth]{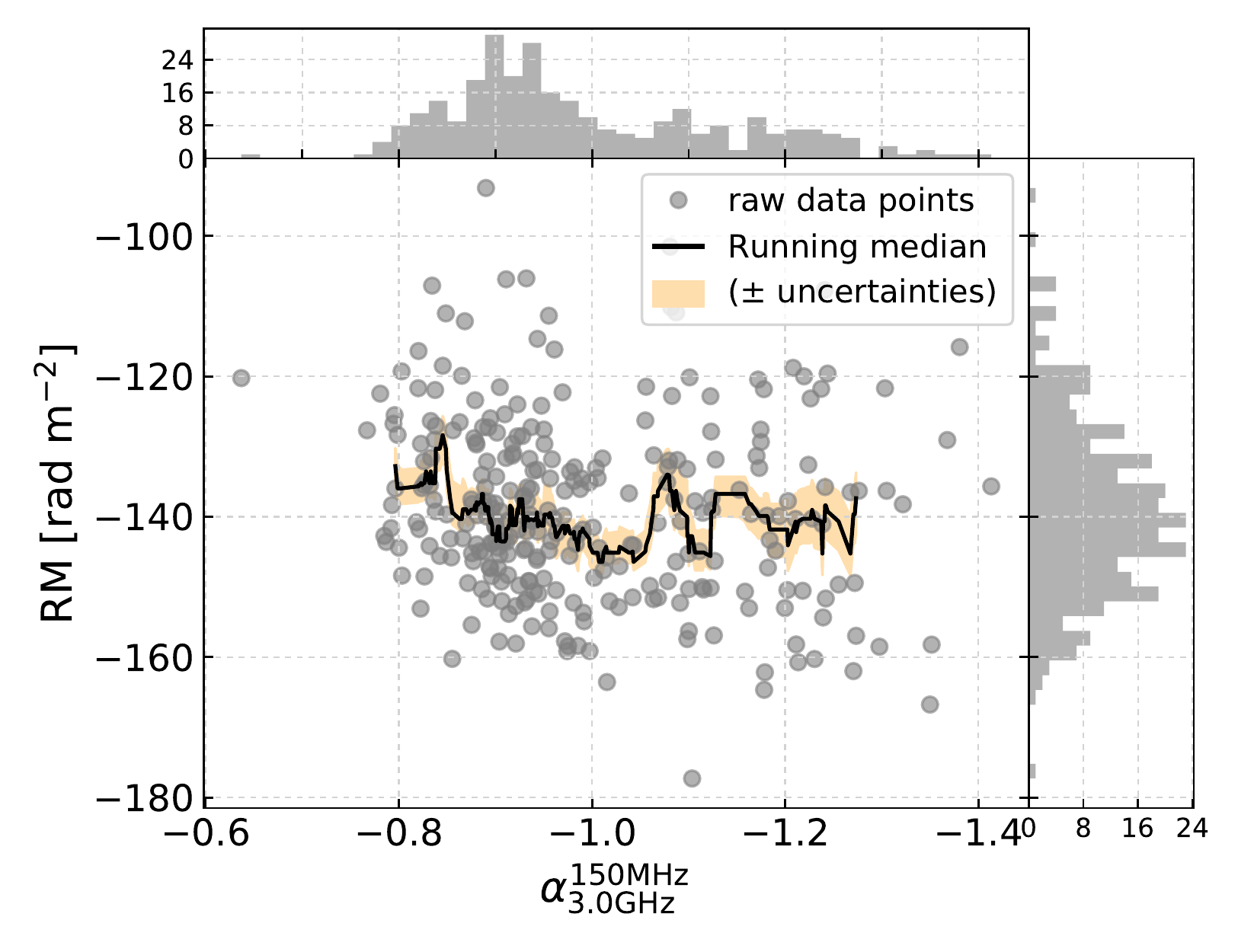}}
\caption{Distributions of the absolute relative Rotation Measure as a function of the spectral index (grey circles). The grey histograms show the projected distribution of the $y$- and $x$-axis quantities along each axis. The black solid line shows the running median of RM in the $\alpha^{\rm 150MHz}_{\rm 3.0GHz}$ space using 20 windows. The yellow area represents the uncertainties on the running median.}
\label{fig:RMvschi_alpha}
\end{figure}

\begin{figure}
\centering
\includegraphics[width=0.45\textwidth]{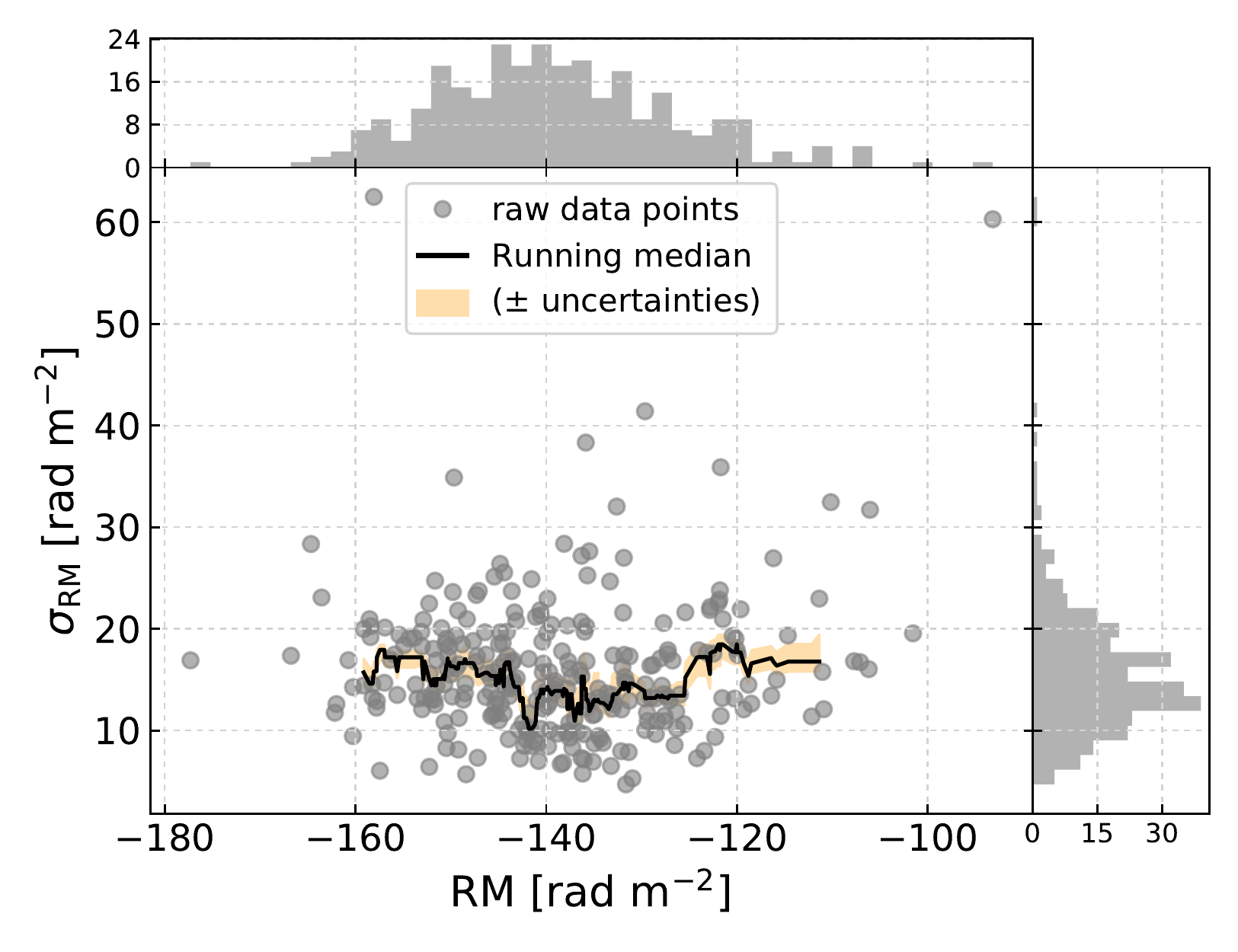}
\caption{Distribution of the external wavelength-dependent depolarization as a function of the absolute relative Rotation Measure (grey circles). The grey histograms show the projected distribution of the $y$- and $x$-axis quantities along each axis. The black solid line shows the running median of $\sigma_{\rm RM}$ in the  RM space calculated using 20 windows. The yellow area represents the uncertainties on the running median.}
\label{fig:RMvsSigmaRM}
\end{figure}

Interestingly, if we extract the profiles of the polarization parameters using an Internal Faraday Rotation Dispersion model (i.e. Eq. \ref{eq:eq:int_far}), we found consistent $p_0$, $\chi_0$ and RM profiles as those we found using the External Depolarization model, and a larger amount of internal depolarization $\varsigma_{\rm RM}$, in agreement with the mathematical differences of the two formulas.  
This means that, with the current data in hand, we cannot distinguish between an External or Internal depolarization model for the northern relic in CIZAJ2242. 
Lower-wavelength wide-band observations (i.e. C- and X-band, 4--8 and 8--12 GHz respectively) might then help to infer the nature of the polarized emission of the northern relic in CIZAJ2242.

\subsection{Investigation for intrinsic RM fluctuations}\label{sect:RMfluctuations}
We found  very weak/no correlations between RM and the spectral index and between RM and the external wavelength-dependent depolarization (Figs. \ref{fig:RMvschi_alpha} and \ref{fig:RMvsSigmaRM}, respectively). The absence of correlation in the latter case is expected in case of external beam depolarization \citep{govoni+04}.

In Sect. \ref{sec:galacticRM}, we show evidence for strong Rotation Measure variation of the Galactic foreground, over angular scales of $3'-5'$, by investigating the RM values in radio galaxies outside the cluster. Along the northern relic, a variation of 30 rad m$^{-2}$ around the median value of 140.8 rad m$^{-2}$ is also found on much smaller scales (i.e. $15''-30''$, see Fig. \ref{fig:RNpolarization}). At the cluster position ($l=104^\circ$ and $b=-5^\circ$), strong variation from the Galactic plane is expected (van Eck, priv. comm.), although detailed studies are still missing. If the detected RM variation is entirely due to the Galactic plane, this would show for the first time that Galactic RM variation is also present on relatively small scales.

Alternatively, this variation could be due to the ICM, and to the magnetic field close to the relic. As shown in Figs. \ref{fig:RNpolarization} and \ref{fig:maps}, the strongest RM fluctuations are measured at the connection of two pairs of filaments, i.e. RN1--RN2 and RN3--RN4, where we measure on average $\Delta{\rm RM}\sim30$ rad m$^{-2}$ (see panel (e) in Fig. \ref{fig:RNpolarization}). 
If this is entirely due to the ICM, given the relation between RM and $B_\parallel$ (Eq. \ref{eq:rm_fit}), we can constrain the magnetic field variation in the relic, being  
$\Delta B_\parallel\sim1~\mu$Gauss, where we have used $n_e=10^{-4}$ cm$^{-3}$ and $L=350$ kpc. Assuming a global value of 5 $\mu$Gauss \citep{vanweeren+10}, we obtain a magnetic field variation of roughly 20\%. In case of weaker global magnetic field, i.e. 1.2 $\mu$Gauss \citep{vanweeren+10}, variations increase up to 80\%.

\begin{table}
\caption{Pearson ($r_p$) and Spearman ($r_s$) rank correlation coefficients of the running median in Figs. \ref{fig:RMvschi_alpha} and \ref{fig:RMvsSigmaRM}.}
\begin{center}
\begin{tabular}{lcc}
\hline
\hline
Parameters & $r_p$ & $r_s$ \\
\hline	
RM--$\alpha^{\rm 150MHz}_{\rm 3.0GHz}$ & $-0.14$ & $-0.17$ \\
$\sigma_{\rm RM}$--RM & $0.08$ & $-0.05$ \\
\hline
\end{tabular}
\end{center}
\label{tab:spearm_coeff_DeltaRM}
\end{table}

\section{Conclusions}\label{sec:conc}

In this work, we have presented a polarimetric study of the merging galaxy cluster CIZA\,J2242.8+5301 ($z=0.1921$) in the 1--4 GHz frequency range with the Jansky Very Large Array. 
We used the $QU$-fitting approach to obtain information on the polarization parameters, i.e. intrinsic polarization fraction ($p_0)$, intrinsic polarization angle ($\chi_0)$, Rotation Measure (RM) and depolarization ($\sigma_{\rm RM}$), for the full cluster at $2.7''$, $4.5''$, $7''$ and $13''$ resolution. This work mainly focused on the most prominent source in CIZA\,J2242.8+5301, i.e., the northern radio relic (RN). Below, we summarize the main results of our work:

\begin{itemize}
\item  CIZA\,J2242.8+5301 is bright in polarized light, with the emission coming from several sources, both diffuse and associated with radio galaxies. In particular, at the highest resolution available (i.e. $2.7''$) the northern relic mimics the filamentary structure seen in total intensity emission \citep{digennaro+18}.

\item In agreement with previous studies \citep{vanweeren+10,kierdorf+17}, we found a high degree of intrinsic polarization in RN, with the eastern side having a higher value than the western one (i.e. $p_{\rm 0, east}\sim 0.55$ and $p_{\rm 0, west}\sim 0.35$, with $p_0$ the best-fit values from the $QU$-fit). 

\item The polarization vectors strongly align with the shock surface also in high resolution observation (i.e. $2.7''$), implying that the magnetic field is dominantly tangled on scales smaller than $\sim8$ kpc.

\item For the first time we were able to investigate the polarization parameters in the relic post-shock region on ten-kpc scales. We found that both the best-fit intrinsic and 1.5 GHz polarization fractions (i.e. $p_0$ and $p_{\rm 1.5GHz}$) decrease towards the cluster center. While, for the latter, a strong contribution of the external wavelength-dependent depolarization is present, the downstream depolarization profile for $p_0$ does not correlate with RM and $\sigma_{\rm RM}$. 

\item We speculate that complex geometrical projections and/or relic shape could possibly explain the $p_0$ downstream depolarization, although detailed modelings should be further worked. We also note that the decrease of the degree of magnetic field anisotropies (i.e. $B_{\rm ord}/B_{\rm rand}$) by about 40\% might explain the depolarization.

\item We detect only one polarized background radio galaxy, i.e. source O. Its $\sigma_{\rm RM}$ is similar to the average value in the northern relic, and allows us to set an approximate value on the turbulent cluster magnetic field of about $5.6~\mu$Gauss.

\item Different Rotation Measures are observed in the northern and southern relics ($\rm RM_{RN}\sim-140$ and $\rm RM_{RS}\sim-80$ $\rm rad~m^{-2}$, respectively). This could be either due to variation of the foreground Galactic Faraday Rotation or to a different contribution of $n_eB_\parallel$ in the ICM along the line of sight. 

\item Rotation Measure fluctuations of about 30 rad m$^{-2}$ on physical scales of about $3'-5'$ are observed at the location of the northern relic. With the current data in hand we cannot determine whether this is due to Galactic plane or to  magnetic field local to the relic. In the former case, this will be the first evidence of small-scale Galactic RM fluctuations. In the latter case, we estimate a magnetic field variation of about 1 $\mu$Gauss.

\end{itemize}

Recently, the polarization properties of radio relics were investigated by \cite{wittor+19} and \cite{roh+19} using numerical simulations. Although they were able to reproduce some properties of observed relics, such as the global observed degree of polarization, they found that it is difficult to explain the high degree polarization (up to $\sim 60$ \%) and the uniformity of the intrinsic polarization angle of the Sausage relic. Incorporating realistic modelings, as well as matching the spatial resolution for simulations and observations, would be crucial steps for the understanding of the observed polarization properties of relics and the connection to the underlying magnetic field.

\begin{acknowledgements}

We thank the anonymous referee for useful comments which have improved the quality of the manuscript. 
GDG and RJvW acknowledge support from the ERC Starting Grant ClusterWeb 804208. HJAR acknowledge support from the ERC Advanced Investigator programme NewClusters 321271. RJvW acknowledges support of the VIDI research programme with project number 639.042.729, which is financed by the Netherlands Organisation for Scientific Research (NWO).
Partial support for LR comes from U.S. National Science Foundation grant AST 17-14205 to the University of Minnesota. DR acknowledges support from the National Research Foundation of Korea through grants 2016R1A5A1013277 and 2020R1A2C2102800. AS acknowledges support through a Clay Fellowship administered by the
Smithsonian Astrophysical Observatory. WF, CJ and RPK acknowledge support from the Smithsonian Institution and the Chandra High Resolution Camera Project through NASA contract NAS8-03060.
This research made use of APLpy, an open-source plotting package for Python \citep{robitaille+bressert12}.

\end{acknowledgements}

\appendix
\renewcommand\thefigure{\thesection.\arabic{figure}}

\section{$QU$-fit plots}\label{apx:dwnpix}
\setcounter{figure}{0} 
In Fig. \ref{fig:fitpixelexample} we show an example of the $QU$-fitting results on a single pixel with high S/N at the shock location. In Fig. \ref{fig:fitpixeldwnexample}, we show the same results but applied on a pixel in the relic downstream. Despite the lower S/N, a single-RM component $QU$ fit still provides a good match to our data. In Figs. \ref{fig:RMSynth}, we show the Faraday spectrum on these two pixels, obtained with \texttt{pyrmsynth}. The RM cube ranges from $-4000$ to $+4000$ rad m$^{-2}$, with a FWHM of 60 rad m$^{-2}$. The two symmetric side-lobes we see next to each peak are likely due to interference in the Faraday spectra, as we do not use the \texttt{RM-CLEAN} option  \citep[see footnote 2 in][]{brentjens11}.

\begin{figure*}
\centering
{\includegraphics[width=\textwidth]{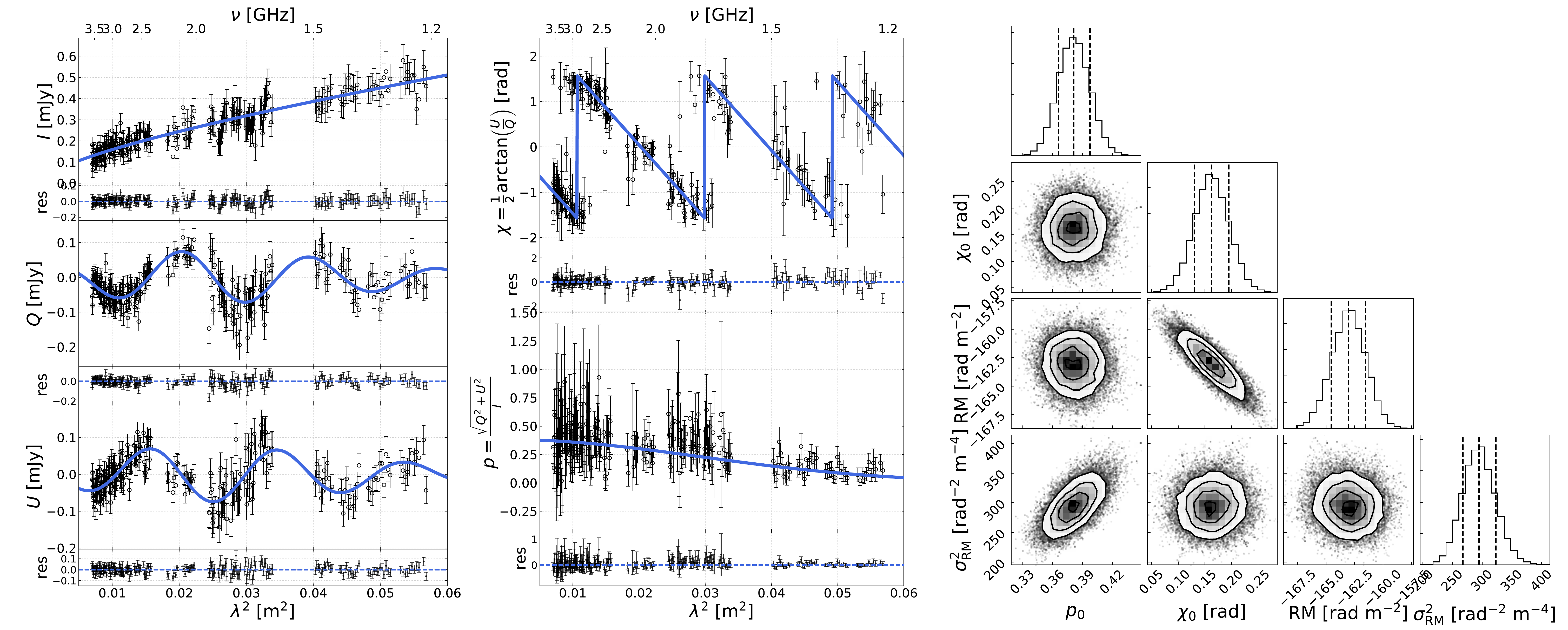}}
\caption{As Fig. \ref{fig:fitpixelexample} but for a pixel further in the RN downstream region.}
\label{fig:fitpixeldwnexample}
\end{figure*}

\begin{figure*}
\centering
{\includegraphics[width=0.3\textwidth]{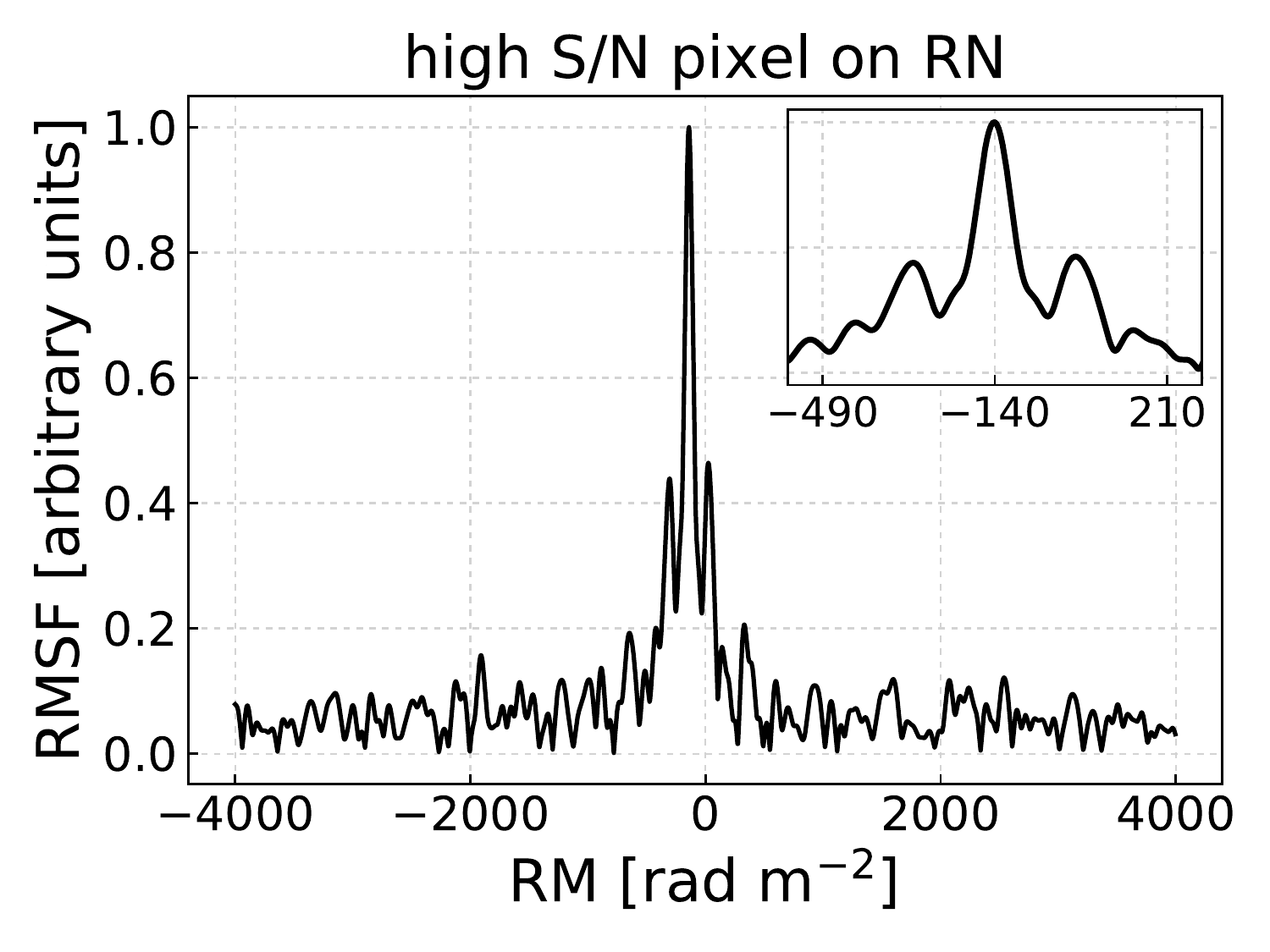}}
{\includegraphics[width=0.3\textwidth]{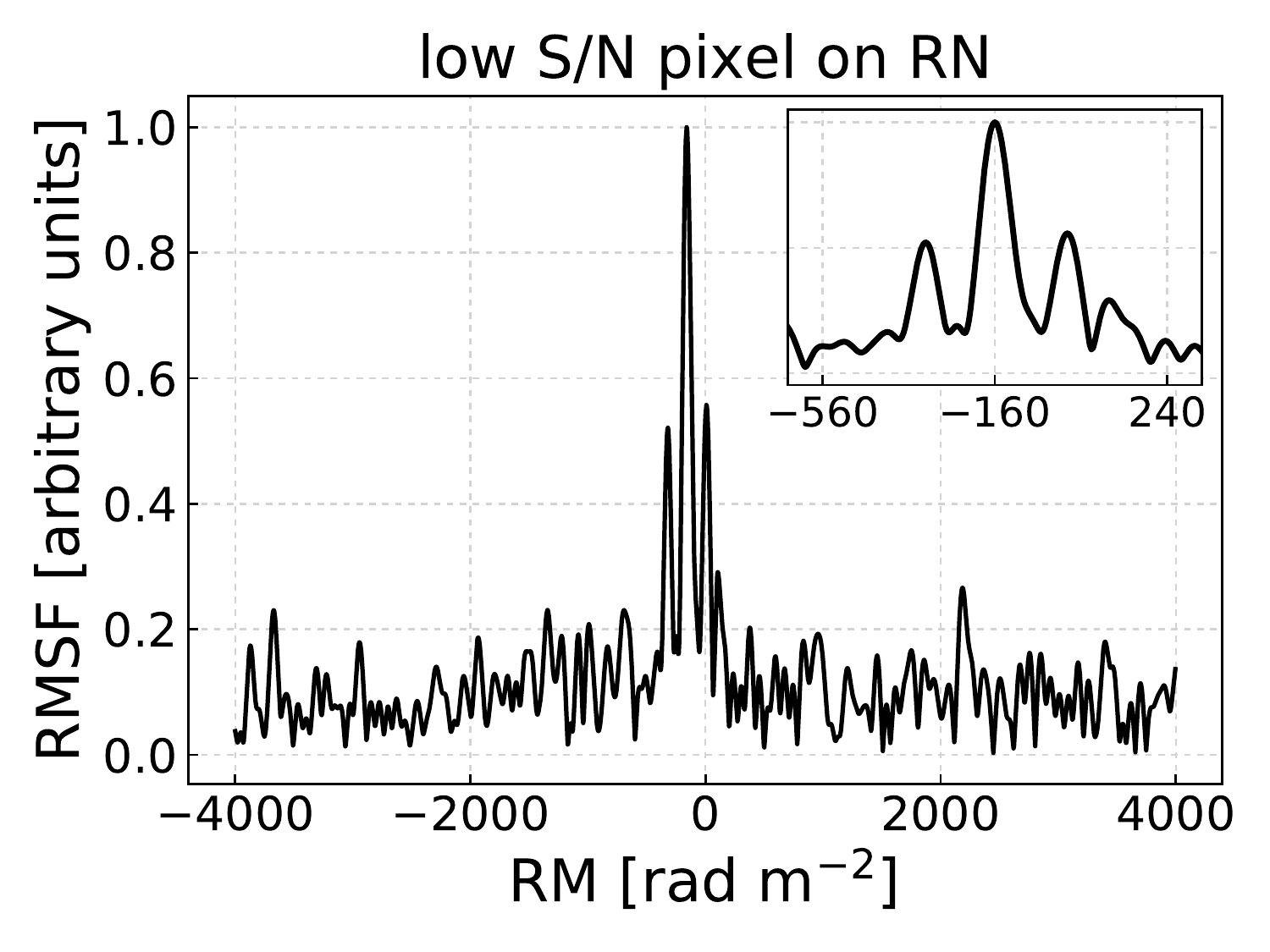}}
{\includegraphics[width=0.3\textwidth]{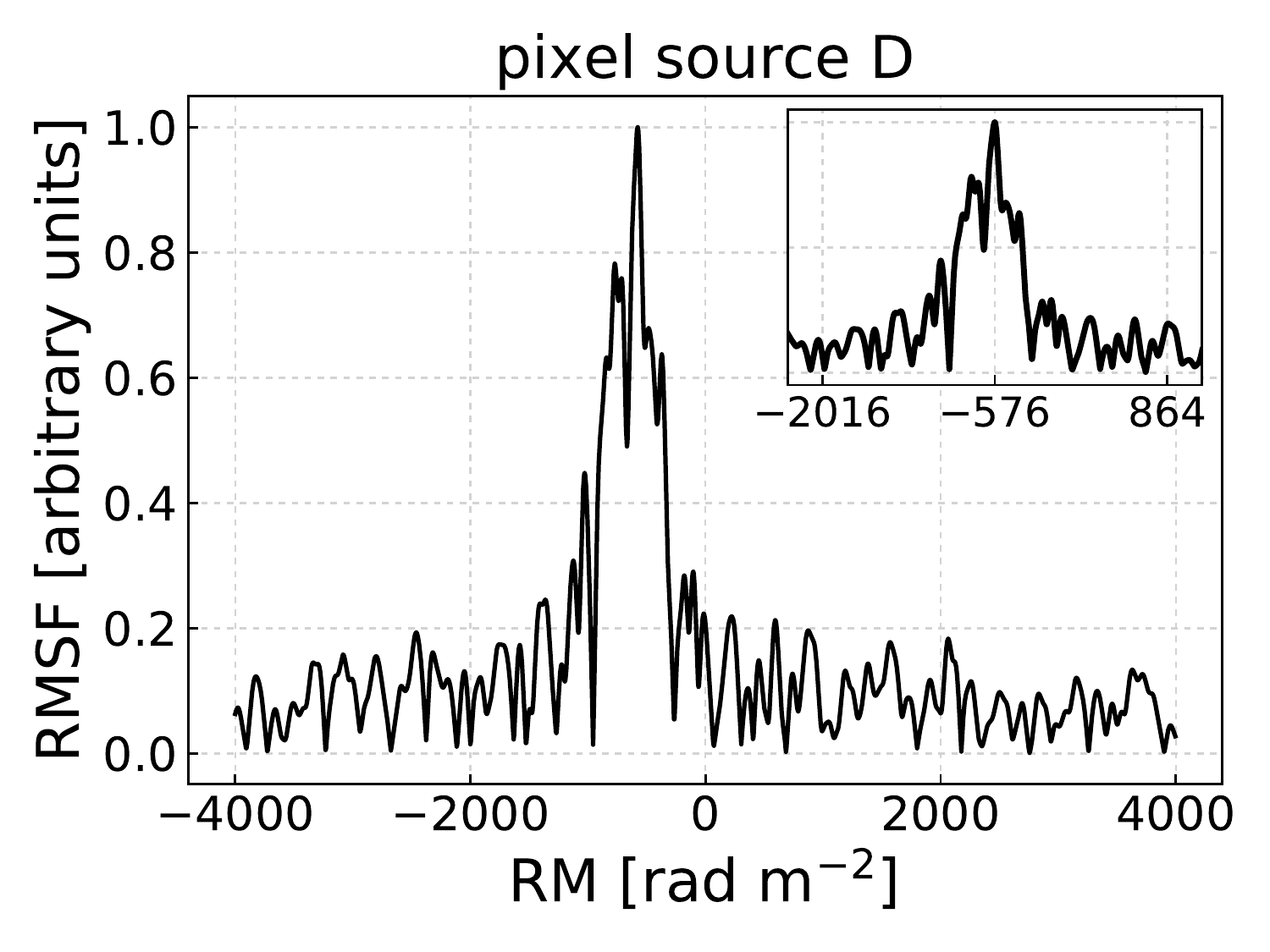}}
\caption{Faraday spectrum on the pixels displayed in Figs. \ref{fig:fitpixelexample} (left panel) and \ref{fig:fitpixeldwnexample} (central panel). In the right panel, the Faraday spectrum of a high S/N pixel in source D is shown. The inset in the two plots shows the zoom on the Faraday peak.}
\label{fig:RMSynth}
\end{figure*}

\section{Uncertainty maps on the polarization parameters}\label{apx:additonal}
\setcounter{figure}{0} 
In this section, we show the $p_0$, RM and $\sigma_{\rm RM}$ negative and positive uncertainty maps correspondent to Fig. \ref{fig:RNpolarization}(d), (e) and (f) (right and left column in \ref{fig:RNmaps_err}), and the $p_{\rm 1.5 GHz}$ uncertainty maps (Fig. \ref{fig:p1.5map_err}). We also present the polarization parameter uncertainty (negative and positive) maps of the full cluster at $13''$ resolution (Figs. \ref{fig:maps_errneg} and \ref{fig:maps_errpos}). The map of the polarization fraction at 1.5 GHz and its correspondent uncertainty map of the full cluster at $7''$ resolution is displayed in Fig. \ref{fig:p1.5map}).

\begin{figure*}
\centering
{\includegraphics[width=\textwidth]{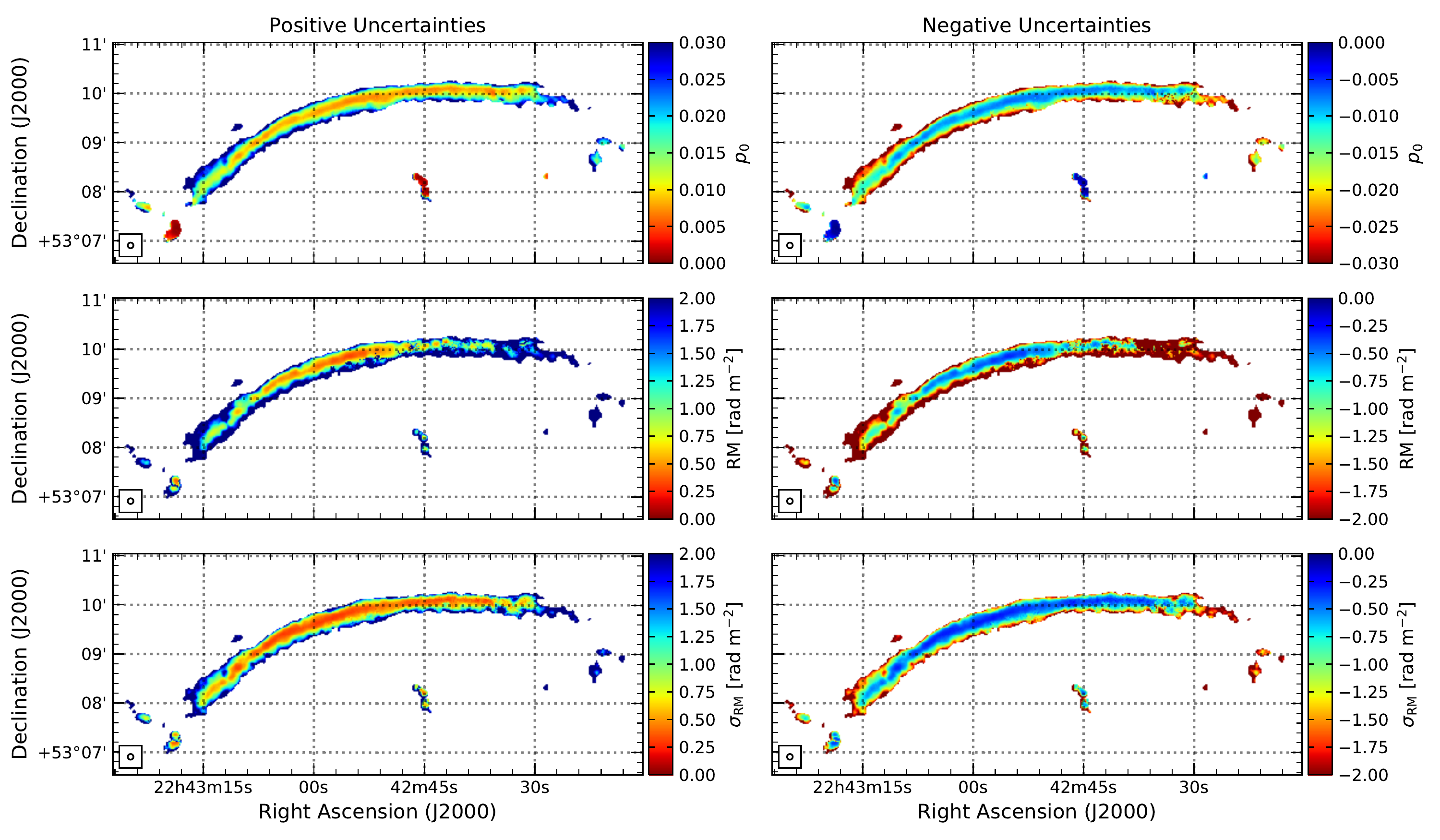}}
\caption{Positive (left column) and negative (right column) uncertainty maps corresponding to panels (c), (e) and (f) in Fig. \ref{fig:RNpolarization}.}
\label{fig:RNmaps_err}
\end{figure*}

\begin{figure}
\centering
\includegraphics[width=0.8\textwidth]{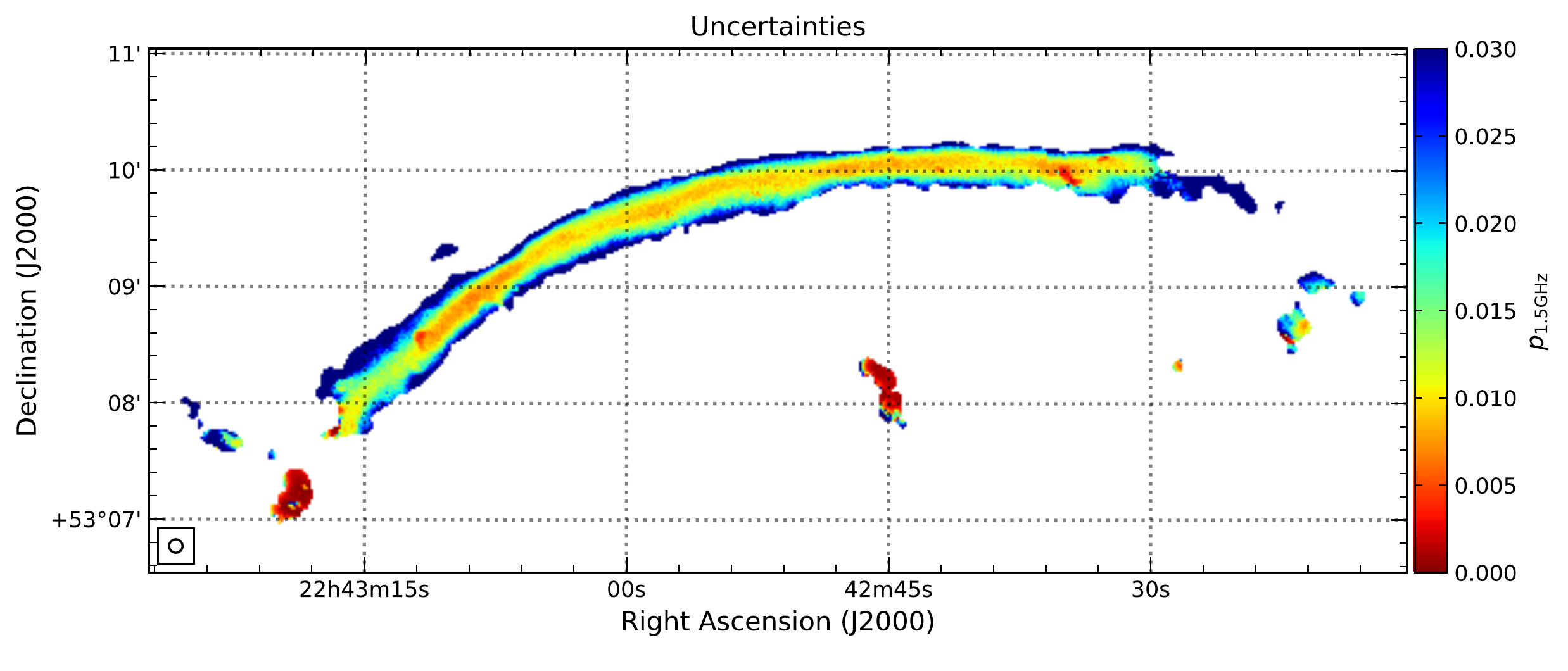}
\caption{1.5 GHz polarization fraction uncertainty map (panel (d) in Fig. \ref{fig:RNpolarization}).}
\label{fig:p1.5map_err}
\end{figure}

\begin{figure*}
\centering
{\includegraphics[width=\textwidth]{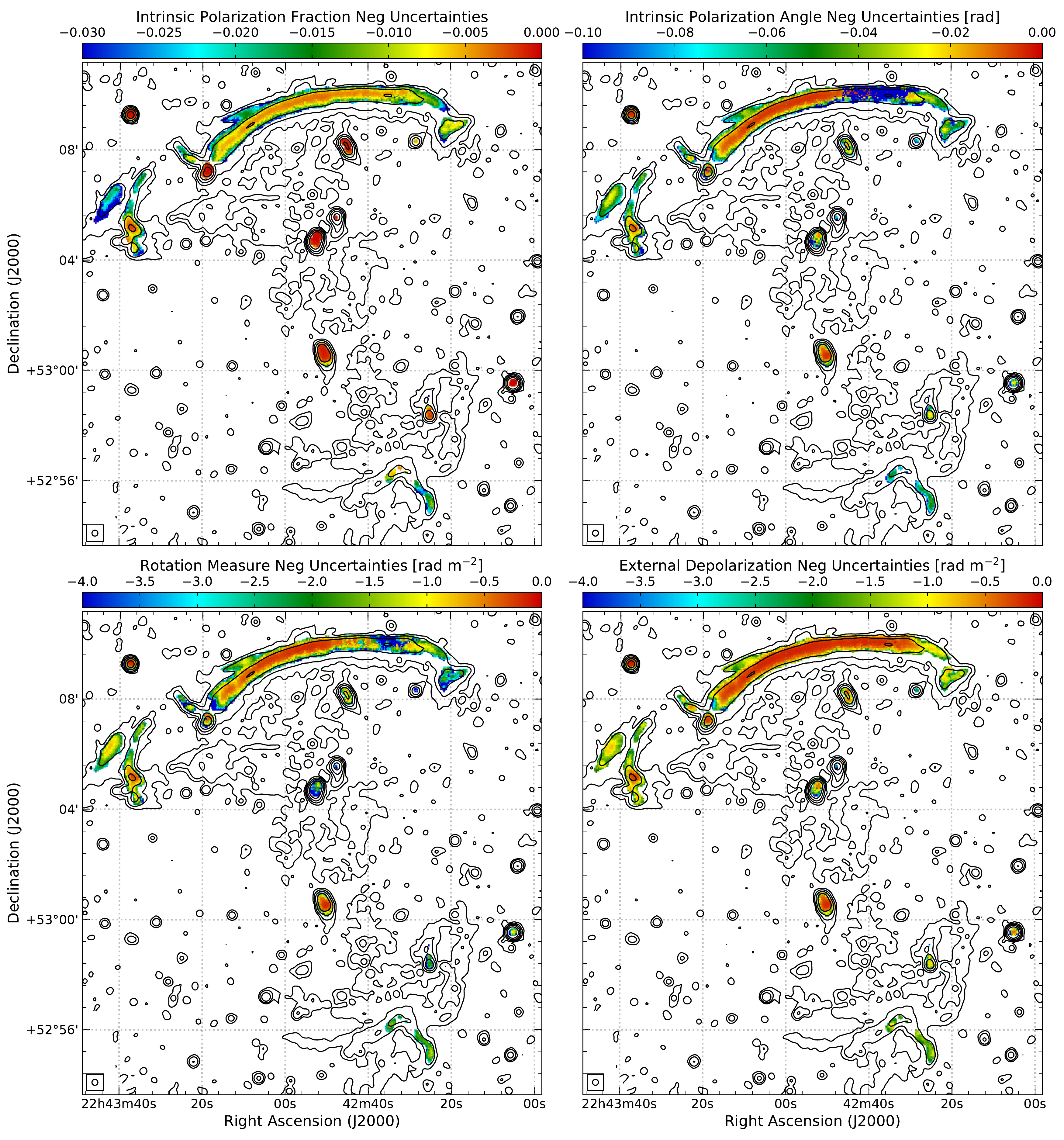}}
\caption{The negative uncertainty maps corresponding to Fig. \ref{fig:maps}.}
\label{fig:maps_errneg}
\end{figure*}

\begin{figure*}
\centering
{\includegraphics[width=\textwidth]{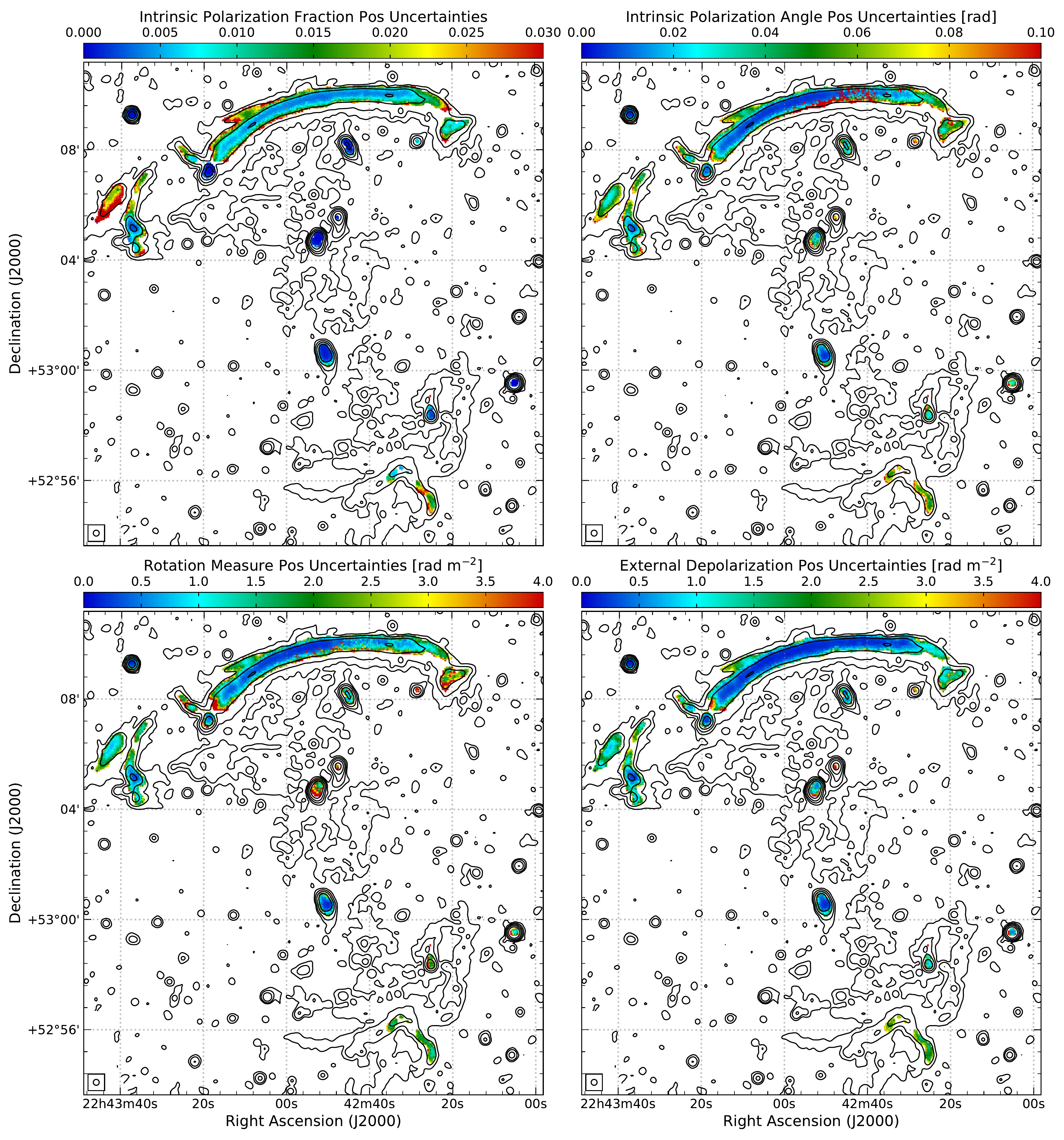}}
\caption{The positive (bottom panel) uncertainty maps corresponding to Fig. \ref{fig:maps}.}
\label{fig:maps_errpos}
\end{figure*}

\begin{figure}
\centering
{\includegraphics[width=\textwidth]{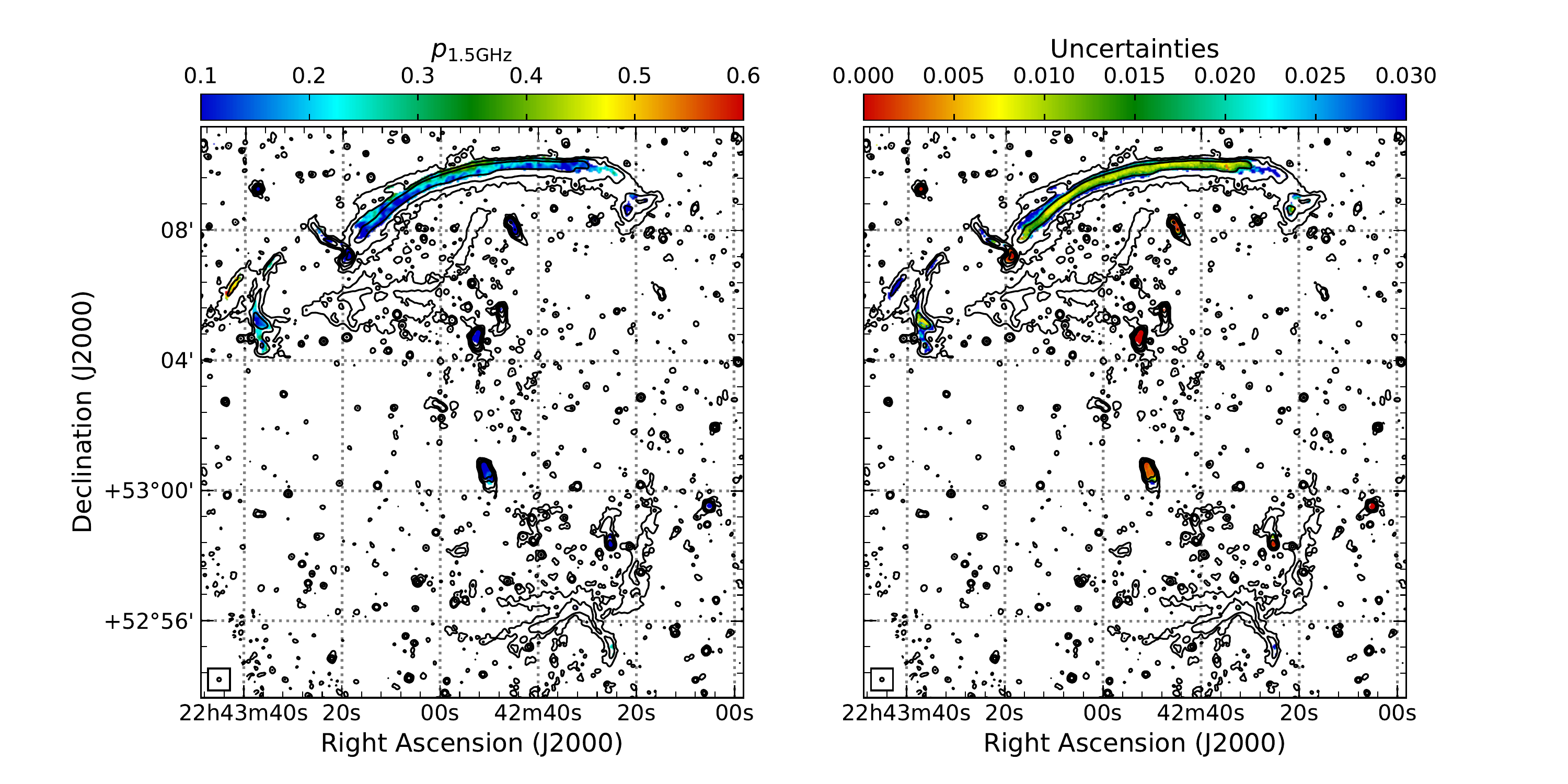}}
\caption{Polarization fraction map at 1.5 GHz (left panel) and correspondent error map (right panel) of CIZAJ2242 at $7''$ resolution. Stokes I radio contours at the same resolution are drawn in black at level of $3\sigma_{\rm rms}\sqrt{1,4,16,64,\dots}$, with $\sigma_{\rm rms}=4.2~\mu$Jy beam$^{-1}$ \citep{digennaro+18}.}
\label{fig:p1.5map}
\end{figure}

\section{Annuli on RN3 and grid used for the correlation analysis}\label{apx:grid}
\setcounter{figure}{0} 
Here, we display the regions where we performed the $QU$-fit. The boxes shown in Fig. \ref{fig:boxesRN3} generate the profiles in Figures \ref{fig:profiles} and \ref{fig:profiles_1.5}. The boxes shown in Fig. \ref{fig:boxes} generate Figures \ref{fig:p0_spix}, \ref{fig:corr_p0_p1.5}, \ref{fig:RMvschi_alpha} and \ref{fig:RMvsSigmaRM}. 
Each box has the same size of the restoring beam, i.e. $7''\times7''$ (about $22\times22$ kpc$^2$ at the cluster redshift). The polarized flux in each box is above a threshold of $3\sigma_{{\rm rms,}P}$ (see Sect. \ref{sec:lambda-fit}).

\begin{figure}
\centering
\includegraphics[width=0.98\textwidth]{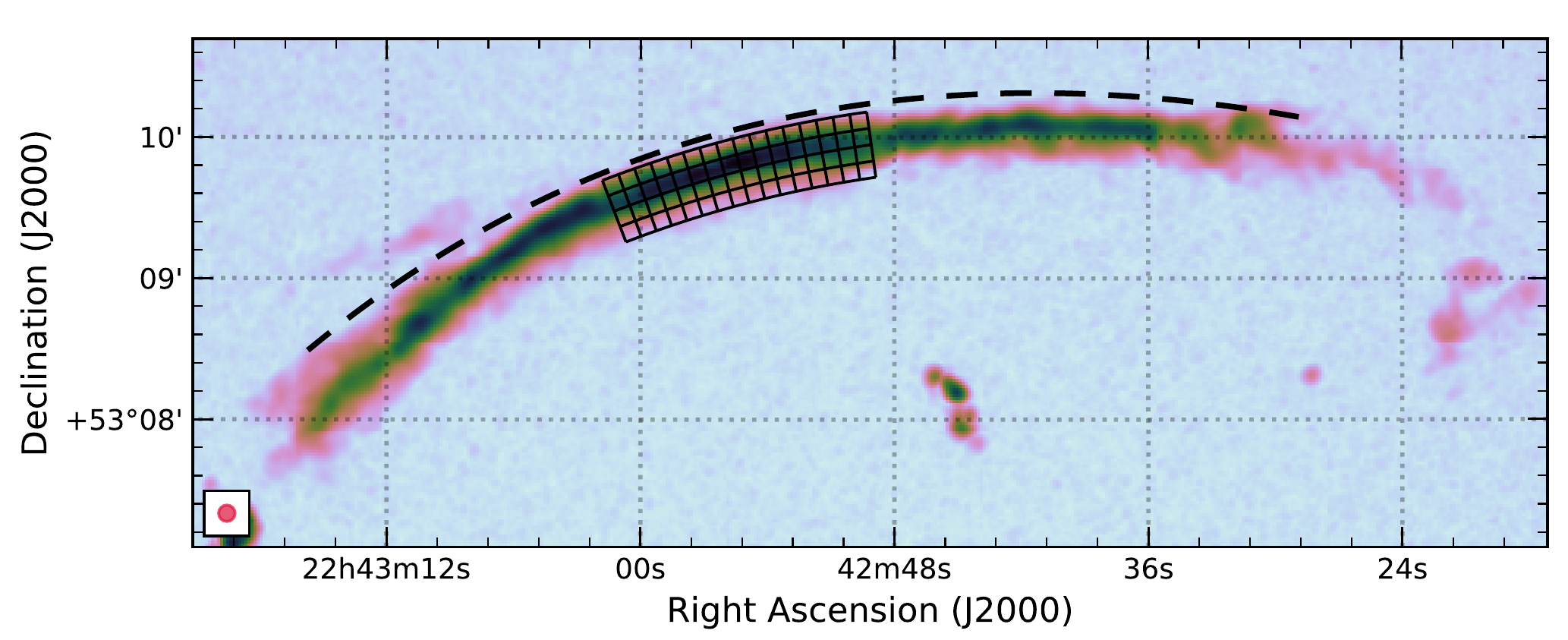}
\caption{Total averaged polarization image at $7''$ resolution of the northern relic with the boxes used to investigate the presence correlation among the polarization parameters in Figs. \ref{fig:profiles} and \ref{fig:profiles_1.5}. The position of the shock (i.e. $d_{\rm shock}=0$ kpc) is displayed by the black dashed line.}
\label{fig:boxesRN3}
\end{figure}

\begin{figure}
\centering
\includegraphics[width=0.98\textwidth]{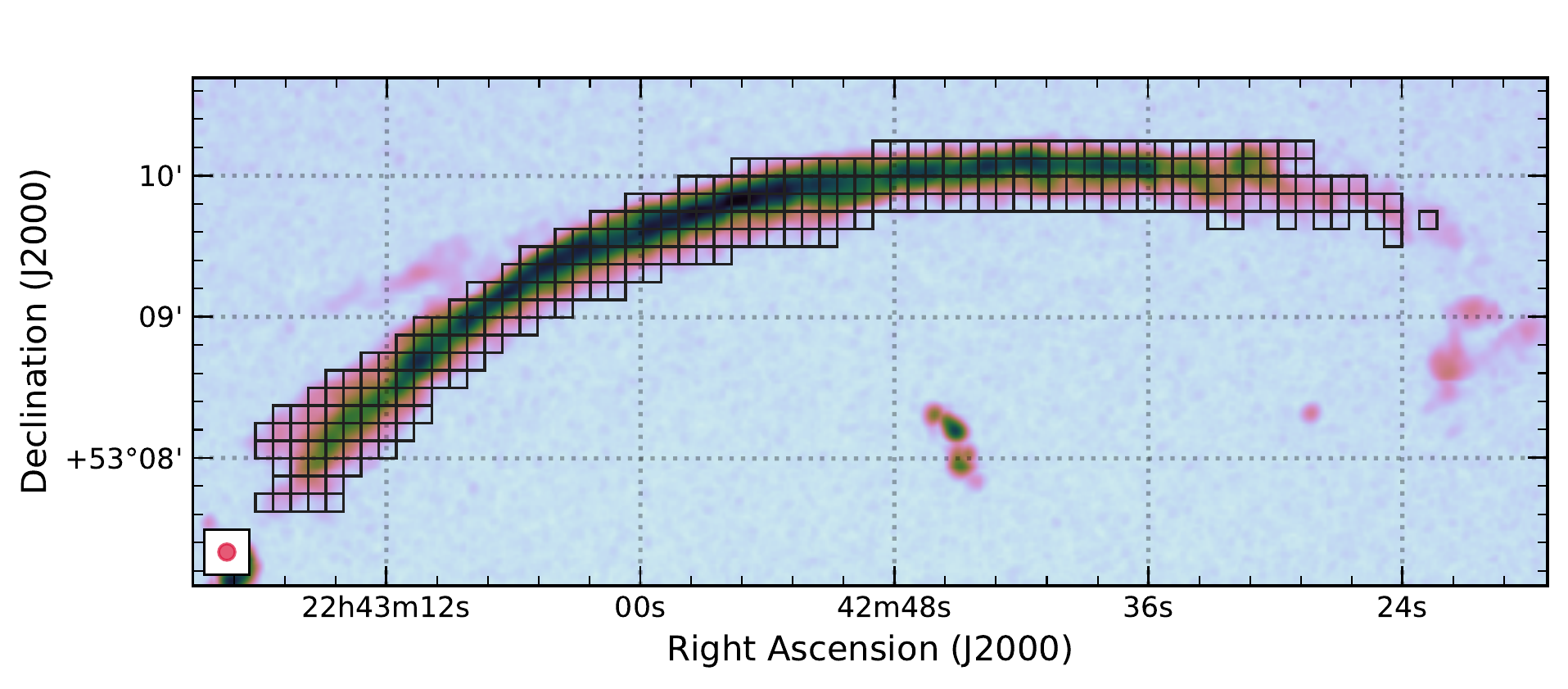}
\caption{Total averaged polarization image at $7''$ resolution of the northern relic with the boxes used to investigate the presence correlation among the polarization parameters in Figs. \ref{fig:corr_p0_p1.5} and \ref{fig:RMvsSigmaRM}.}
\label{fig:boxes}
\end{figure}

\bibliography{biblio.bib}

\end{document}